\documentclass[fleqn,usenatbib]{mnras}

\usepackage{newtxtext,newtxmath}

\usepackage[T1]{fontenc}

\DeclareRobustCommand{\VAN}[3]{#2}
\let\VANthebibliography\thebibliography
\def\thebibliography{\DeclareRobustCommand{\VAN}[3]{##3}\VANthebibliography}
\usepackage{graphicx}	
\usepackage{amsmath}	
\usepackage{caption}
\usepackage{subcaption}
\usepackage{multicol}

\title[Interacting Galaxies in IllustrisTNG -- VIII]{Interacting galaxies in the IllustrisTNG simulations -- VIII: Pericentric star formation rate enhancements are driven both by increased fuelling and efficiency.}

\author[L. Faria et al.]{
Lawrence Faria,$^{1,2}$\thanks{E-mail: lawrence.faria@queensu.ca}
David R. Patton,$^{2}$
Stéphane Courteau$^{1}$
Sara Ellison$^{3}$
and Westley Brown$^{4,2}$
\\
$^{1}$Department of Physics and Astronomy, Queen's University, Kingston, Ontario K7L 3N6, CA\\
$^{2}$Department of Physics and Astronomy, Trent University, 1600 West Bank Drive, Peterborough, ON K9L 0G2, Canada\\
$^{3}$Department of Physics and Astronomy, University of Victoria, Finnerty Road, Victoria, BC, V8P 1A1, Canada\\
$^{4}$Department of Physics and Astronomy, York University, 4700 Keele Street, Toronto, ON M3J 1P3, Canada
}

\date{Accepted XXX. Received YYY; in original form ZZZ}

\pubyear{2024}

\begin{document}
\label{firstpage}
\pagerange{\pageref{firstpage}--\pageref{lastpage}}
\maketitle

\begin{abstract}
Using the TNG100-1 cosmological simulations, we explore how galaxy properties, such as specific star formation rate ($\rm sSFR=SFR/M_*$), gas fraction ($\rm f_{gas} \,= \, M_{\rm H}/M_{*}$), and star formation efficiency ($\rm SFE_{H} = SFR/M_{H}$), change over the course of galaxy-galaxy interactions.
We identify 18,534 distinct encounters from the reconstructed orbits of a sample of massive galaxies ($\rm M_{*} > 10^{10} \; \rm M_{\odot}$) with companions within a stellar mass ratio of 0.1 to 10.
Using these encounters, we study the variation of galaxy properties over time as they approach and move away from pericentric encounters over a redshift range of $0 \leq z < 1$.
Following the closest pericentric encounters ($\leq 50$ kpc) of a host galaxy with its companion, we find that sSFR is enhanced by a factor of $1.6 \pm 0.1$ on average within the central stellar half-mass radius (R\textsubscript{1/2}) compared to pre-encounter values.
Our results show a time delay between pericentre and maximum sSFR enhancement of $\sim$0.1 Gyr with a mean galaxy separation of 75 kpc.
We similarly find that $\rm f_{gas}$ is enhanced by a factor of $1.2 \pm 0.1$, and $\rm SFE_{H}$ is enhanced by a factor of $1.4 \pm 0.1$ following the pericentre of an encounter within the same timescale.
Additionally, we find evidence of inflowing gas towards the centre, measured by comparing the $\rm f_{gas}$ and metallicity within the central R\textsubscript{1/2} to the galactic outskirts.
We find that approximately 70 per cent of the peak sSFR enhancement can be attributed to the increase in $\rm SFE_{H}$, with the increase in $\rm f_{gas}$ contributing the remaining 30 per cent.

\end{abstract}




\section{Introduction}
Interactions and mergers of galaxies play a pivotal role in the formation and evolution of galaxies within the $\Lambda$CDM paradigm.
Modern observations and simulations have found that mergers and interactions significantly perturb the properties and morphologies of galaxies. 
Mergers and interactions trigger enhancements in the star formation rate \citep{BartonGellerKenyon2000, Ellison2008, Woods2010, Patton2013, Ellison2020, Bickley2022}, trigger active galactic nuclei activities (AGNs; \citealt{Alonso2007, Ellison2011, Satyapal2014, Ellison2019, Bickley2023, Byrne-Mamahit2023, Bickley2024, Byrne-Mamahit2024}), dilute gas-phase metallicities \citep{Ellison2008, Kewley2010, Rupke2010, Scudder2012a, Bustamante2020}, potentially completely transform their morphology and structure \citep{HernadezToledo2005, Patton2005, DePropris2007, Casteels2014, Patton2016} and quench star formation \citep{Ellison2022, Li2023, Ellison2024}.
The degree to which these properties deviate from those of isolated galaxies depends on the orbital dynamics of the interaction \citep{Moreno2015, Garduno2021}, disk spin orientations of the galaxies \citep{DiMatteo2007, Moreno2015, Vasiliev2022}, gas content and star formation rate prior to the interaction \citep{Scudder2015, Pan2018}, mass ratio of the galaxy pair \citep{Woods2006, Ellison2008, Lambas2012, Pan2018} and their environment \citep{Ellison2010, Scudder2012b, Moreno2013, Contreras-Santos2022}.

The observed effects of galaxy-galaxy interactions on gas and star formation are typically explained by a combination of the enhancement of the molecular gas reservoir and an increase in the efficiency of star formation \citep{Kaneko2017, Ellison2018, Pan2018, Violino2018}. 
High-resolution merger simulations \citep{MihosHernquist1996, Hopkins2008,Torrey2012, Moreno2015, Kaneko2017, Moreno2019, Vasiliev2022} suggest that interactions and mergers create strong gravitational torques within interacting galaxies, triggering gas inflows towards the central regions.
Inflowing gas would result in the observed centrally diluted metallicity \citep{Ellison2008, Lin2010, Torrey2012, Scudder2012a}. 
Furthermore, accretion of gas (due to interactions/mergers) onto the central supermassive black hole may further trigger AGNs \citep{Ellison2011, Byrne-Mamahit2023, Byrne-Mamahit2024}.

In addition to this fresh supply of gas, interacting galaxies are also shown to form stars more efficiently \citep{Saintonge2012, Michiyama2016, Saintonge2017, Violino2018}.
Observations of star-forming galaxies find that while the surface density of star formation rate typically scales with that of the molecular gas, as predicted by the Kennicutt-Schmidt relation, offsets from this trend are driven by variations in the star formation efficiency (SFE; \citealt{Utomo2017, Silverman2018, Ellison2020b}).  
It remains an open question whether the enhancements in star formation observed in interacting galaxies are primarily driven by an enhanced fuel supply or enhanced SFE \citep{Pan2018, Moreno2021, Thorp2022, Sargent2024}.  

Galaxy pair interactions and mergers occur on the timescale of $\sim$1 billion years \citep{Patton2002, Lotz2008}. 
As a result, observations of interacting systems can only happen at discrete epochs and not as a continuous time-series measurement.
Star formation rate (SFR) is typically enhanced in pairs at small projected separations (\textless 30 kpc), with the pre-merger enhancement being strongest in the closest pairs \citep{Ellison2008, Li2008}. 
However, \citet{Patton2013} in their study of galaxy pairs from the Sloan Digital Sky Survey (SDSS) found compelling evidence that SFR enhancements can be observed out to separations as large as 150 kpc. 
Merger simulations find that enhancements in SFR are typically largest in a galaxy following the first pericentric passage of its companion and then again at the time of merger \citep{Springel2005, Cox2006, Cox2008, Lotz2008, Torrey2012, Hopkins2013}.
Large enhancements in SFR, which are more easily detectable, appear short-lived while smaller enhancements may endure for longer periods of time \citep{DiMatteo2007}.
As such, the conclusions drawn from studies utilising pairs catalogues may be significantly coloured by the epoch at which the pair is observed. 
Additionally, interacting galaxy pairs within observations are identified through the presence of stellar debris fields, disturbed discs, tidal features, or the presence of a close neighbour \citep{KitzbichlerWhite2008}. 

Studies have attempted to group galaxy pairs into interaction stages based on separation and morphology \citep{Larson2016, Pan2019}, but this is understood to be limited in its accuracy, especially for post-encounter pairs at wide separations \citep{Patton2016} as well as flybys and direct mergers.
This leads to studies of galaxy pairs naturally taking the form of linking properties changed by an interaction to the projected separation of the galaxy pair, rather than to a specific pericentre of the interaction.
This limits our ability to fully understand the physical processes that occur over the course of an interaction and which give rise to the changes observed.
Comparing the properties of interacting galaxies to those of relatively isolated galaxies is an additional complexity. 
With the diversity in galaxy formation, to what extent can we accurately match one galaxy to another, especially when the properties of the interacting galaxies observed may have already been significantly affected by an encounter? 
Indeed, control selection matched on a variety of galaxy properties such as SFR, stellar mass, pair separation, gas content, or environment can affect the conclusions of pair catalogue studies \citep{Scudder2015, Patton2016, Pan2018}. 

Merger simulations bridge this gap by allowing us to model the full orbit of a single galaxy pair from first in-fall to eventual merger in high mass and time resolution \citep{Springel2003, MihosHernquist1996, Hopkins2008, Torrey2012, Moreno2019}. 
However, due to the computational intensity, these simulations are only able to model an interaction with pre-selected orbital parameters for galaxies with a limited variation of pre-encounter properties.
Additionally, these simulations do not typically model the low-density circumgalactic medium or wider environment in which such interaction takes place. 
Galaxy pairs are observed in a wide variety of environments and the degree to which their properties are affected by an interaction has also been shown to have some dependence on that environment \citep{Barton2007, Ellison2010, Scudder2012b, Contreras-Santos2022}.
As such, high-resolution merger simulations, though an extremely valuable tool for better understanding the internal mechanisms of interacting galaxies, lack the required cosmological context to directly link their results to observations. 
Cosmological hydrodynamical simulations such as Illustris \citep{Vogelsberger2014}, Horizon-AGN \citep{Dubois2014}, EAGLE \citep{Schaye2015}, SIMBA \citep{Dave2019} and IllustrisTNG \citep{Nelson2019}, despite their lower temporal and spatial resolution, provide a large, diverse group of galaxies, in diverse environments, with properties consistent with observed galaxies at present day. 

In the first paper in this series, \citet{Patton2020} constructed a large catalogue of galaxy pairs in the IllustrisTNG simulation. 
\citet{Patton2020} endeavoured to better connect the results of high-resolution merger simulations to the observed properties of galaxy pairs using this simulated large statistical cosmological sample. 
They effectively used an observational method of studying galaxy pairs relative to separation on this large sample of simulated galaxies. 
\citet{Patton2020} found that the specific star formation rates (sSFR) of host galaxies are enhanced by a factor of 2 at small separations with the companion galaxy, with observable enhancements out to separations of 200 kpc. 
This result was found to be consistent with observations when repeated on a sample of galaxy pairs from the Sloan Digital Sky Survey \citep{Abazajian2009}.
\citet{Brown2023}, a subsequent paper in this series, confirms that these findings are consistent whether the host galaxy is actively star-forming or quiescent.
Additionally, other papers in this series have investigated star formation rates and AGN activity in paired and post-merger galaxies within the IllustrisTNG simulation
\citep{Hani2020, Quai2021, Byrne-Mamahit2023, Byrne-Mamahit2024}.
Building on this, \citet{Patton2024} reconstructed the orbits of the IllustrisTNG galaxy pairs identified in their 2020 study. 
Their primary goals included identifying and characterizing significant past and future close encounters between the galaxies in each pair and determining the time of merger for pairs that subsequently merge. 
These reconstructions provided insights into the frequency of close encounters and mergers and provide us with the unique opportunity to analyse changes in galaxy properties as a function of time.

In this study, we aim to further bridge the gap between merger simulations and observations by using the IllustrisTNG simulation to investigate how galaxy properties change over the course of individual encounters.
In particular, we focus on the sSFR, gas fractions (f\textsubscript{gas}), star formation efficiency of hydrogen gas (SFE\textsubscript{H}), and gas metallicities of galaxies during these interactions. 
By leveraging the reconstructed orbits from \citet{Patton2024}, we can track the variation in these properties as a function of time relative to the pericentre of individual close encounters.
Furthermore, we aim to investigate whether enhanced f\textsubscript{gas} or enhanced SFE\textsubscript{H} acts as the primary driver for enhanced star formation in interacting galaxies.

In the following section, we describe the simulations and the methodology used to identify our dataset of discrete encounters. 
In Section \ref{sec:results}, we present our findings on the variation of sSFR, $f_{\text{gas}}$, and SFE$_{\text{H}}$ across individual encounters, along with the relative contributions of enhanced fuel supply and increased efficiency to the star formation triggered by the encounter.
In Section \ref{sec:disc}, we discuss these findings and attempt to orient them in the current literature on galaxy pairs in observations and simulations.
Finally, in Section \ref{sec:conclusion} we summarise our conclusions and describe future work that will benefit from the methodology and/or the encounters dataset presented in this study. 

\section{Dataset and Methodology}\label{sec:data}
\subsection{IllustrisTNG}
This work uses a dataset assembled from the public release of the IllustrisTNG simulation suite \citep{Nelson2019, Naiman2018,Marinacci2018,Pillepich2018a,Springel2018}. 
IllustrisTNG is a large-scale cosmological simulation that models the formation and evolution of galaxies from the early universe to the present day using the AREPO moving-mesh code \citep{Springel2010}. 
The simulation tracks dark matter, gas, stars, and black holes across three virtual volumes (TNG50, TNG100, TNG300), with data presented in 100 snapshots from redshift z = 127 to z = 0.
Stars are allowed to form stochastically from gas cells following the Kennicutt-Schmidt relation \citep{Kennicutt1998}, which relates the SFR surface density $\Sigma_{\rm SFR}$ to the cold gas surface density $\Sigma_{\rm cold~gas}$. 
Astrophysical processes such as turbulent and magnetic field support in the interstellar medium, supernovae, and star formation are accounted for in the subgrid physics of the simulation.
The conversion of gas cells to stars is adjusted to realistically account for these baryonic processes and produce realistic galaxies \citep{Pillepich2018}.
Galaxy identification in the simulation is performed using the {\sc Subfind} algorithm, which associates particles and cells to subhalos based on gravitational and kinematic linkage. 
However, the algorithm can erroneously associate particles from one subhalo to another, particularly for closely interacting galaxies. 
This issue, known as numerical stripping, is a known limitation which we account for with the selection criteria of our dataset \citep{Rodriguez-Gomez2015}. 
Following \citet{Patton2020} and \citet{Patton2024}, our dataset uses the TNG100-1 run of the IllustrisTNG simulation, for which we have reconstructed orbits.

\subsection{Encounters Dataset}\label{subsec:encount}
The primary goal of this study is to quantify the change in the properties of interacting galaxies during pericentric encounters. 
As such, we aim to track how galactic properties change over the timescale of individual pericentres, comparing the properties before and after the pericentre.
The properties of interacting galaxies in high-resolution merger simulations appear to change before or around any sufficiently close pericentre, with some of these changes persisting long after the encounter \citep{Torrey2012, Moreno2019}.

We begin with \citet{Patton2020}'s catalogue of galaxies and their closest companions from the TNG100-1 run, assembled via the methodology described in \citet{Patton2016} and \citet{Patton2020}. 
Host galaxies are required to be within the stellar mass limits of $10^{10}M_{\odot} < M_{*} < 10^{12}M_{\odot}$. 
Closest companion galaxies are the subhalo with the smallest 3D separation from the host, with a stellar mass of at least 10 per cent the stellar mass of the host galaxy.\footnote{The companion may therefore be more or less massive than its host galaxy.}

This catalogue is limited to snapshots 50 to 99, which corresponds to a redshift range of 1 to 0.
We then use the reconstructed orbits of these galaxy pairs, presented in \citet{Patton2024}, to identify when the pericentre of a galaxy pair interaction occurs as well as the separation of the galaxy pairs at the time of pericentre. 
We can then find the time of each snapshot relative to the pericentre ($\Delta$t).
This allows us to track galaxy properties as a function of $\Delta$t. 

Within the reconstructed orbits of \citet{Patton2024}, there are a variety of encounters emerging from the diverse sample of orbits; however, we specifically aim to study distinct encounters which are reasonably well separated from other pericentres in the orbit.
This allows us to better link observed effects to the pericentre of interest while minimising the risk of contamination from other pericentres.
The situation we aim to avoid is the inability to determine whether a change in a galactic property, observed between two pericentres, is an aftereffect of the previous pericentre or a precursor to the next pericentre.
Thus, from \citet{Patton2024}'s catalogue of reconstructed orbits, we select distinct pericentres to compile our novel encounters dataset. 
In general, a distinct pericentre is defined using the following selection criteria:
\begin{itemize}
  \item At least 1 Gyr ($\approx 6$ snapshots) has elapsed since the previous pericentre or the start of the orbit if there is no previous pericentre.
  \item At least 4 snapshots between the selected pericentre, and either the next pericentre, the merger of the interacting galaxies, or the final available snapshot.
  \item At least 3 consecutive snapshots following the pericentre in question with the separation of the galaxy pair being greater than the separation at their pericentre.
  \item The companion must remain the closest companion of the host galaxy in the snapshots immediately before and after the pericentre in question.
\end{itemize}
Once identified, snapshots around the pericentres are selected to assemble encounters such that there are three pre-pericentre snapshots and at least one post-pericentre snapshot.
There is no limit to the maximum number of post-pericentre snapshots, allowing as many as the above criteria permit for a given interaction. 

Our goal is to first capture each galaxy at a time before the distinct pericentre of the interaction with their companion, where they are unlikely to have already been perturbed by the interaction. 
We then follow each galaxy across the pericentre and after it, tracking how their properties evolve until such a time that changes in their properties can no longer be feasibly linked to the pericentre. 
We aim to track the response of the properties of the galaxies to the pericentres of their encounters with their closest companion. 
While there may be additional companions, we assume that the changes in the galaxies' properties will be most strongly affected by the pericentric passage of their closest companion. 
Moreover, as we wish to study how galaxy pair interactions affect star formation in galaxies, we require all galaxies to have a non-zero specific star formation rate in the first snapshot of the encounter. 
Our analysis treats all pericentre passages equally, as our selection criteria require at least 1 Gyr between successive encounters to minimize the influence of prior passages. 
We also note that 70 per cent of our encounters are first passages within our redshift range of $\rm 0 \leq z < 1$, though this does not account for potential earlier passages outside this redshift.
The pericentre selection criteria above also attempt to further alleviate the effects of numerical stripping by avoiding the final stages in the merger process where \citet{Patton2024}'s interpolation is more uncertain.
An additional cut is made based on the size-mass relation of TNG100-1 galaxies with stellar masses in the range $10^{10}M_{\odot} < M_{*} < 10^{12}M_{\odot}$, as presented in \citet{Genel2018}, to remove any obvious outliers. 
Using the general trend of the size-mass relation for star-forming galaxies (Figure 2(a) of \citet{Genel2018}), we define an approximate upper boundary based on the observed scatter in the data. 
Galaxies with sizes lying well beyond the upper extent of the 16th–84th percentile range around the median radius are excluded. 
Specifically, we identified and excluded 32 galaxies with unphysical sizes ($> 150$ kpc), indicative of numerical stripping.

Finally, we note that we make no restrictions on our dataset based on environment as we aim to study the effects of these close encounters for galaxies which may exist in any variety of environments.
An examination of how galaxy properties may differ for close encounters in different environments in sub-samples of our dataset is left for future work.
We identify 18,534 distinct pericentres in encounters with 3 pre-pericentre snapshots and at least 1 post-pericentre snapshot, yielding a total sample of 145,219 snapshots.

\begin{figure}
    \includegraphics[width=0.5\textwidth]{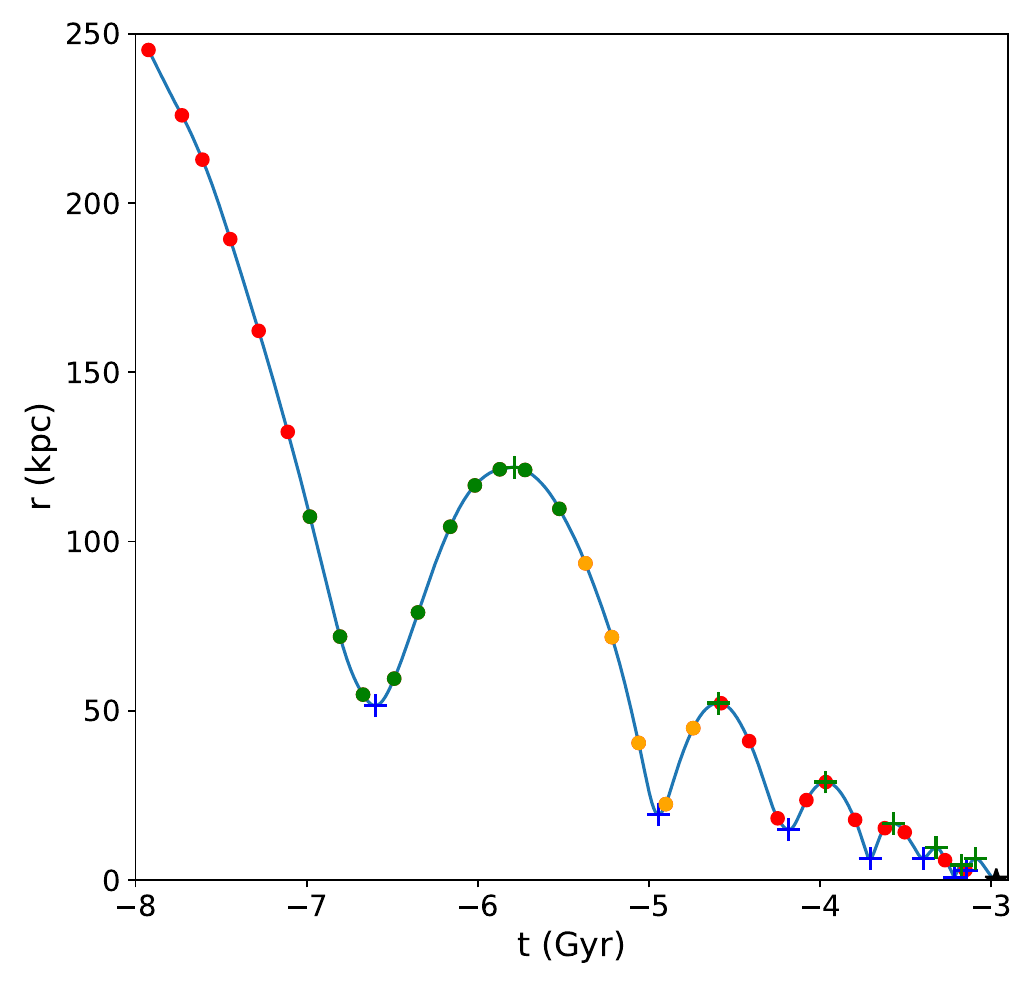}
    \caption[Galaxy Pair Orbit]{3D separation (r) versus time relative to the present day (t) for the reconstructed orbit of a galaxy pair. 
    The red circles represent the data obtained from the consecutive discrete snapshots of TNG100-1, while the blue line represents the results of the 6D kinematic interpolation.
    The blue and green crosses represent identified pericentres and apocentres, respectively. 
    The green and orange circles show snapshots around pericentres which pass our selection criteria for the encounters dataset, whereas the red circles denote snapshots that do not fall within our encounters dataset.}
    \label{fig:orbit}
\end{figure}

Figure \ref{fig:orbit} shows an example of a reconstructed orbit. 
Seven encounters can be seen within this reconstructed orbit; however, only two encounters are sufficiently separated from the other encounters such that they satisfy the criteria for our encounter dataset.
The green and orange circles identify these two encounters. 
As the galaxies move towards their eventual merger, the orbit shrinks and pericentres become more frequent. 
As such, pericentres after the encounter denoted by the orange circles do not pass our selection criteria to be included as encounters.

\begin{figure*}
    \begin{center}
        \includegraphics[width=0.8\textwidth]{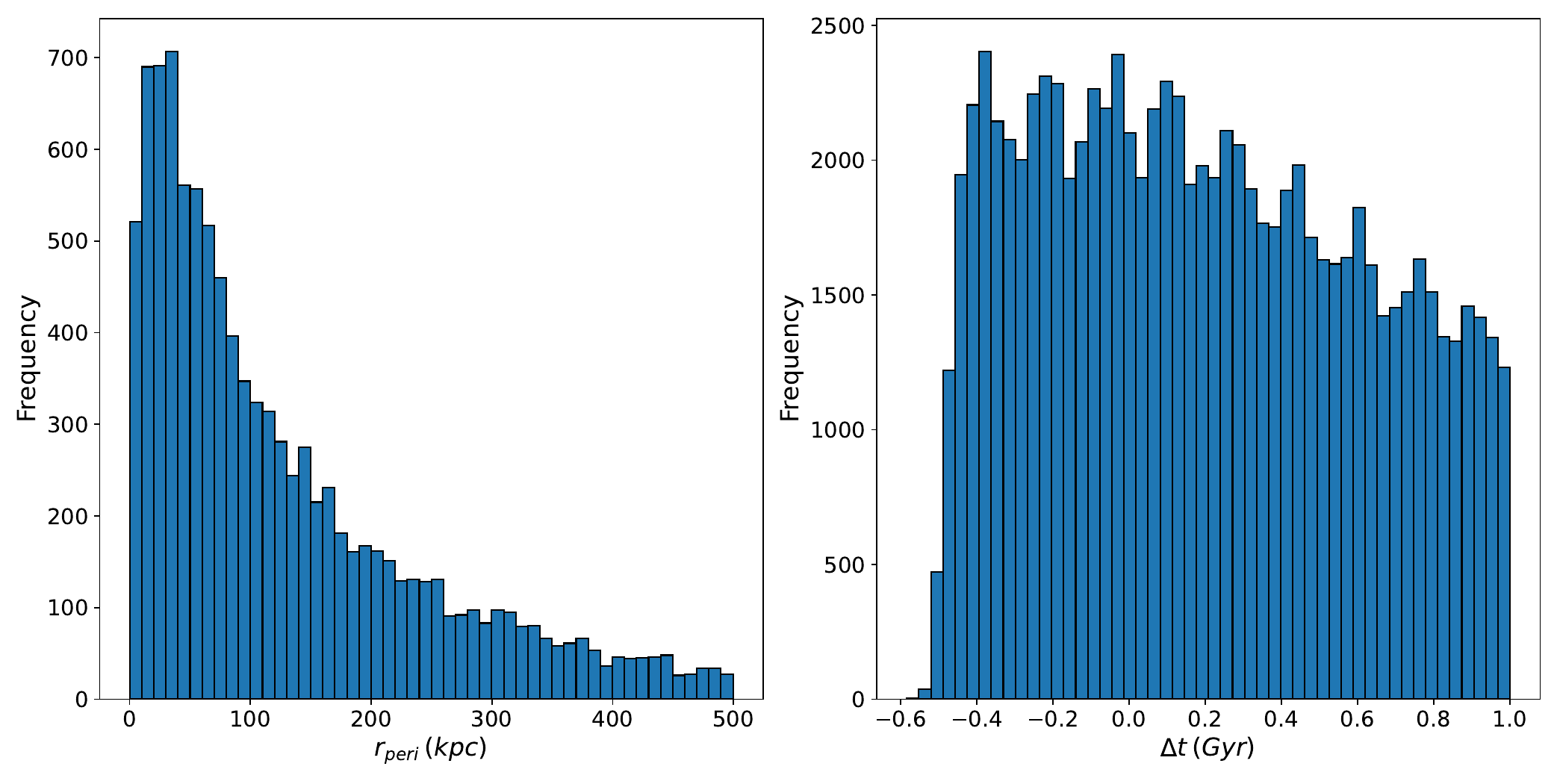}
        \caption[Histograms of r\textsubscript{peri} and $\Delta$t]{Distribution of galaxy pair separation at pericentre, r\textsubscript{peri} (left panel) and time relative to pericentre, $\Delta$t (right panel). The left panel shows the range of r\textsubscript{peri} emergent from our selection criteria ranging from 0 - 500 kpc, while the right panel shows the time ranging from 0.5 Gyr before the pericentre (denoted by a negative $\Delta$t) to 1 Gyr after the pericentre (denoted by a positive $\Delta$t).}
        \label{fig:rperideltathist}
    \end{center}
\end{figure*}

The left panel of Figure \ref{fig:rperideltathist} shows the distribution of the galaxy pair separation at pericentre, r\textsubscript{peri}, in the encounters dataset.
We find that despite the limitations of the orbital interpolation schemes and selection criteria at small r\textsubscript{peri}, we are able to produce a sample with reasonable density at close pericentres.
The r\textsubscript{peri} distribution of the dataset peaks for pairs with pericentre separations of 0-50 kpc, and decreases in number out to separations of 500 kpc, where we impose an upper limit. 
In the right panel of Figure \ref{fig:rperideltathist}, we show the distribution of $\Delta t$ values, which represent the time relative to pericentre in the encounters dataset.
We find that the $\Delta$t values are approximately uniform across the time period of significant interest, from 0.5 Gyr prior to the pericentre to 0.5 Gyr following it, encompassing the typical 1 Gyr time period for a close galaxy pair interaction.
This uniform density in $\Delta$t allows us to examine the average change in the properties of these galaxies continuously in time.
However, we note that there are some fairly regular variations in the distribution of the $\Delta$t values, resulting in an apparent periodicity. 
These variations appear to span 0.15 Gyr, which is close to the approximate length of the interval between IllustrisTNG snapshots. 
When predicting the location of a pericentre between two snapshots through the 6D kinematic interpolation, we ideally anticipate a uniform distribution of predictions across those two points for a dataset of sufficient size. 
However, in our dataset, we see that there is some shape to the distribution.
As we move from snapshot to snapshot, we would expect to see the shape of this distribution repeated, which would create the observed periodicity in Figure \ref{fig:rperideltathist} repeated at time intervals similar to those between two snapshots.
This suggests that the minor repeated variations seen in Figure \ref{fig:rperideltathist} are likely systematic.

\section{Results}\label{sec:results}
With our dataset of encounters compiled according to the criteria outlined in Section \ref{subsec:encount}, we now explore the evolution of galaxy properties throughout these interactions. After defining the key properties in Section \ref{subec:defs}, we analyze their time-dependent behavior in Sections \ref{subsec:sSFR}, \ref{subsec:galprop}, and \ref{sec:metallicty}, and further investigate the primary mechanisms driving star formation triggered by encounters in Section \ref{subsec:fuelfrac}.

\subsection{Definitions}\label{subec:defs}
We examine the specific star formation rate, sSFR, as defined by: 

\begin{equation}
    sSFR = \frac{SFR}{M_{*}}.
\end{equation}

The supply of gas potentially available for star formation in a galaxy is ideally tracked by measuring the mass of cold, dense molecular hydrogen gas ($M_{H_2}$).
Unfortunately, the mass of molecular gas is not reported in the TNG100-1 public data release. 
Therefore, our analysis tracks gas supply via the mass of hydrogen gas, M\textsubscript{H}, which includes all phases of hydrogen gas (atomic, ionized, and molecular). 
We define the gas fraction, $\rm f_{gas}$, as:

\begin{equation}
    \rm f_{gas} = \frac{M_{\rm H}}{M_{*}}.
\end{equation}

The mass of hydrogen gas introduces additional complexity to our definition of SFE, which is commonly defined as $\rm SFE = SFR / M_{H_{2}}$ \citep{Young1996, Boselli2001, Leroy2008}.
Within the simulation, SFE is a combination of the Kennicutt–Schmidt depletion time (which is dependent on $\Sigma_{\rm cold~gas}$) and the mass and density of the gas.
We instead consider SFE\textsubscript{H}, the star formation efficiency dependent on M\textsubscript{H}, the total hydrogen gas in all phases.
This serves as a measure of the efficiency of a galaxy to assemble star-forming gas cells for conversion to stars, as well as the efficiency of converting star-forming gas into stars.
This is defined as:

\begin{equation}\label{eq:SFEHeq}
    SFE_{H} = \frac{SFR}{M_{H_{2}}} \cdot \frac{M_{H_{2}}}{M_{H}}.
\end{equation}

These three quantities can finally be related as follows:

\begin{equation} \label{eq:fineq}
    sSFR =  SFE_{\rm H} \cdot \rm f_{gas}.
\end{equation}
\begin{figure*}
  \centering
  \begin{minipage}{0.45\textwidth}
    \centering
    \includegraphics[width=\textwidth]{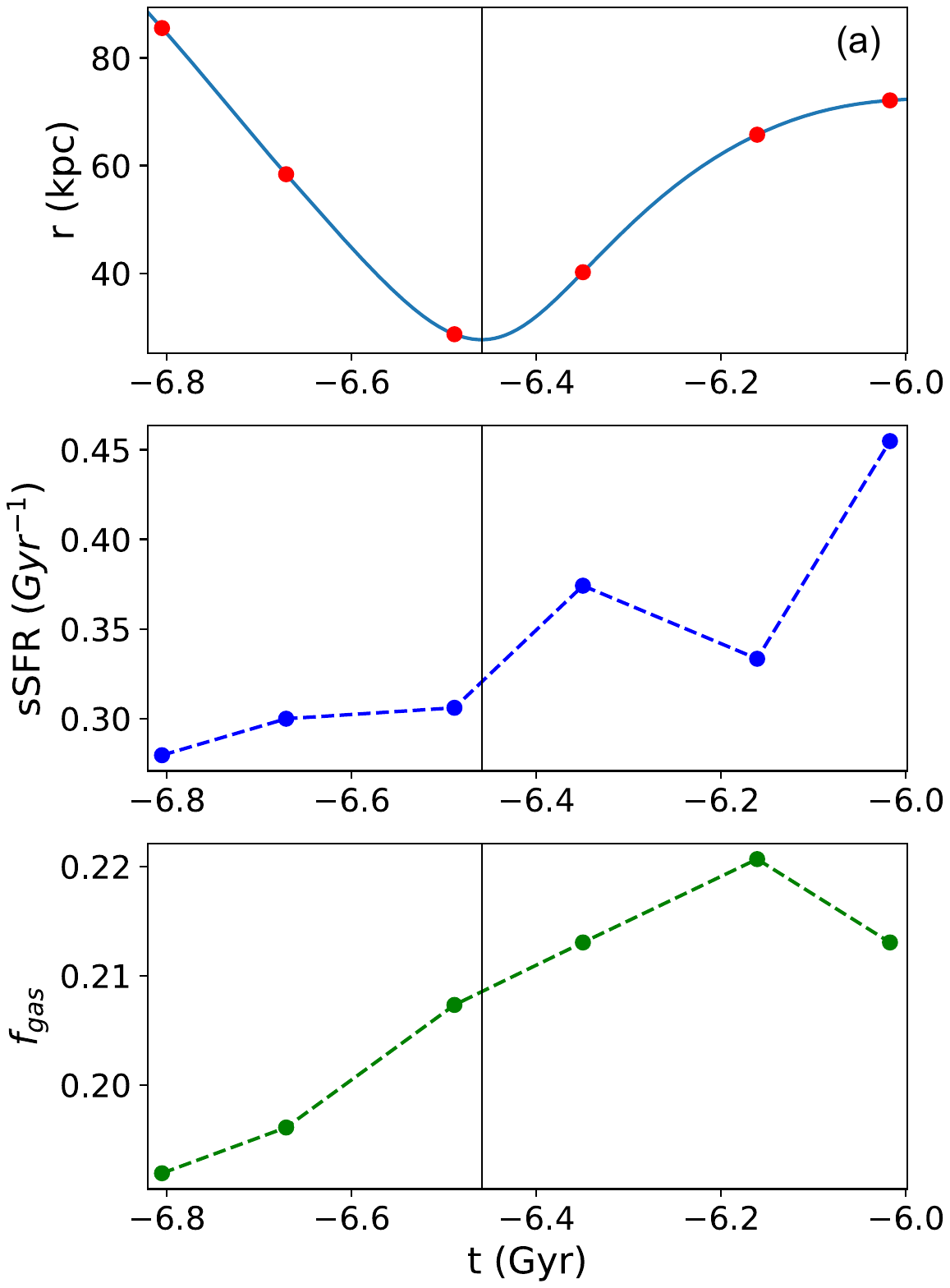}
  \end{minipage}
  \hspace{0.01\textwidth}
  \begin{minipage}{0.45\textwidth}
    \centering
    \includegraphics[width=\textwidth]{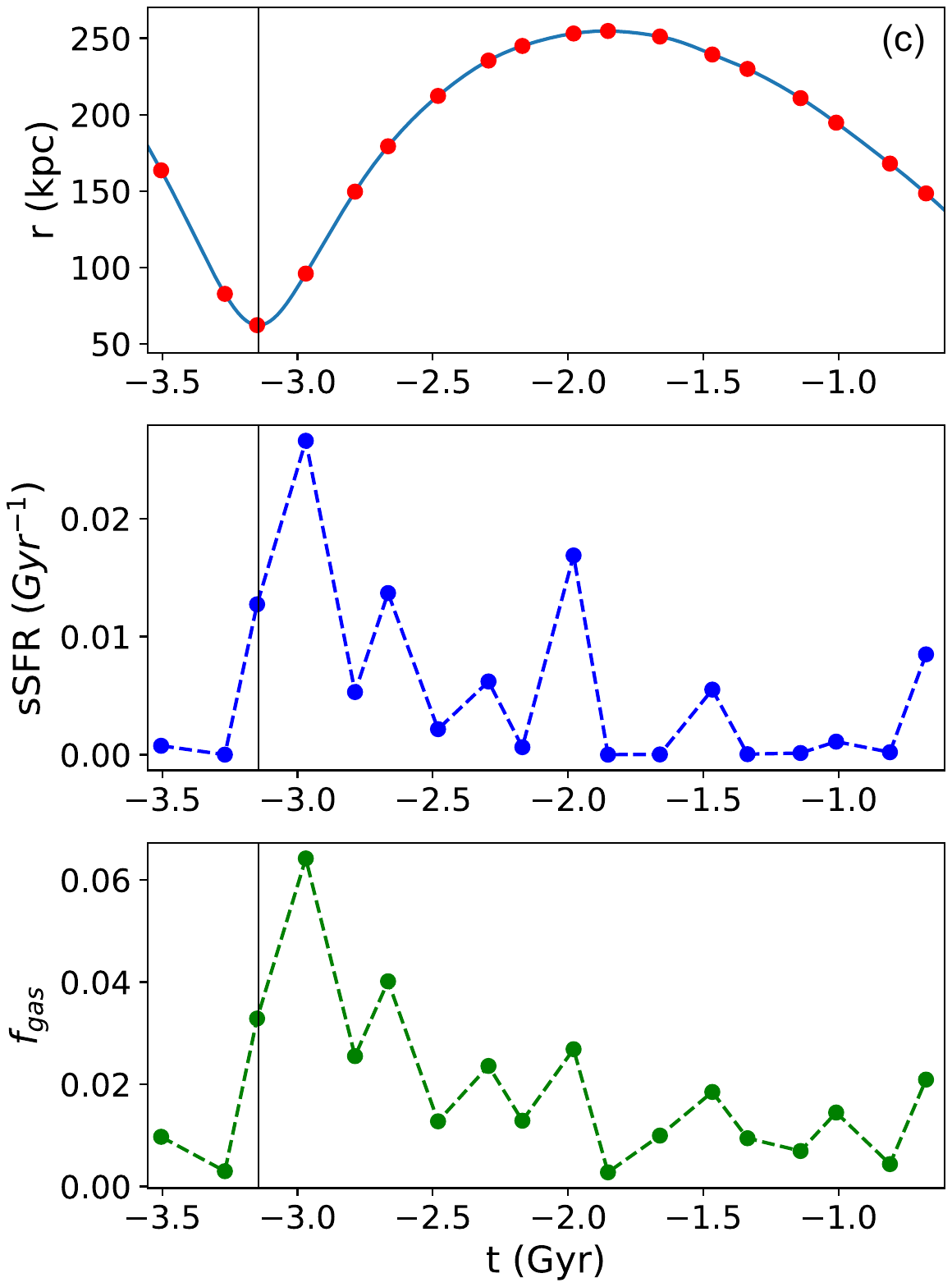}
  \end{minipage}


  \begin{minipage}{0.45\textwidth}
    \centering
    \includegraphics[width=\textwidth]{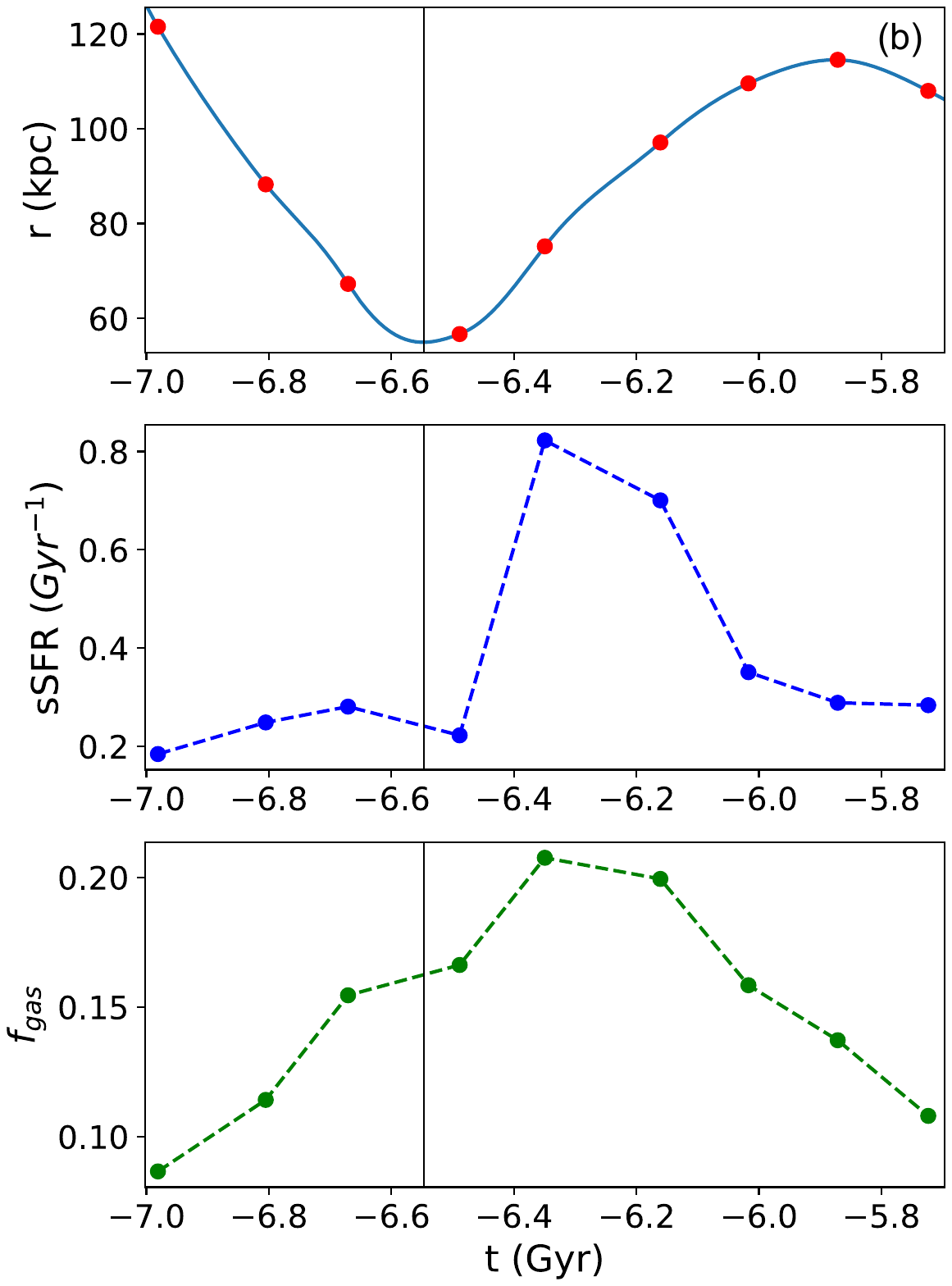}
  \end{minipage}
  \hspace{0.01\textwidth}
  \begin{minipage}{0.45\textwidth}
    \centering
    \includegraphics[width=\textwidth]{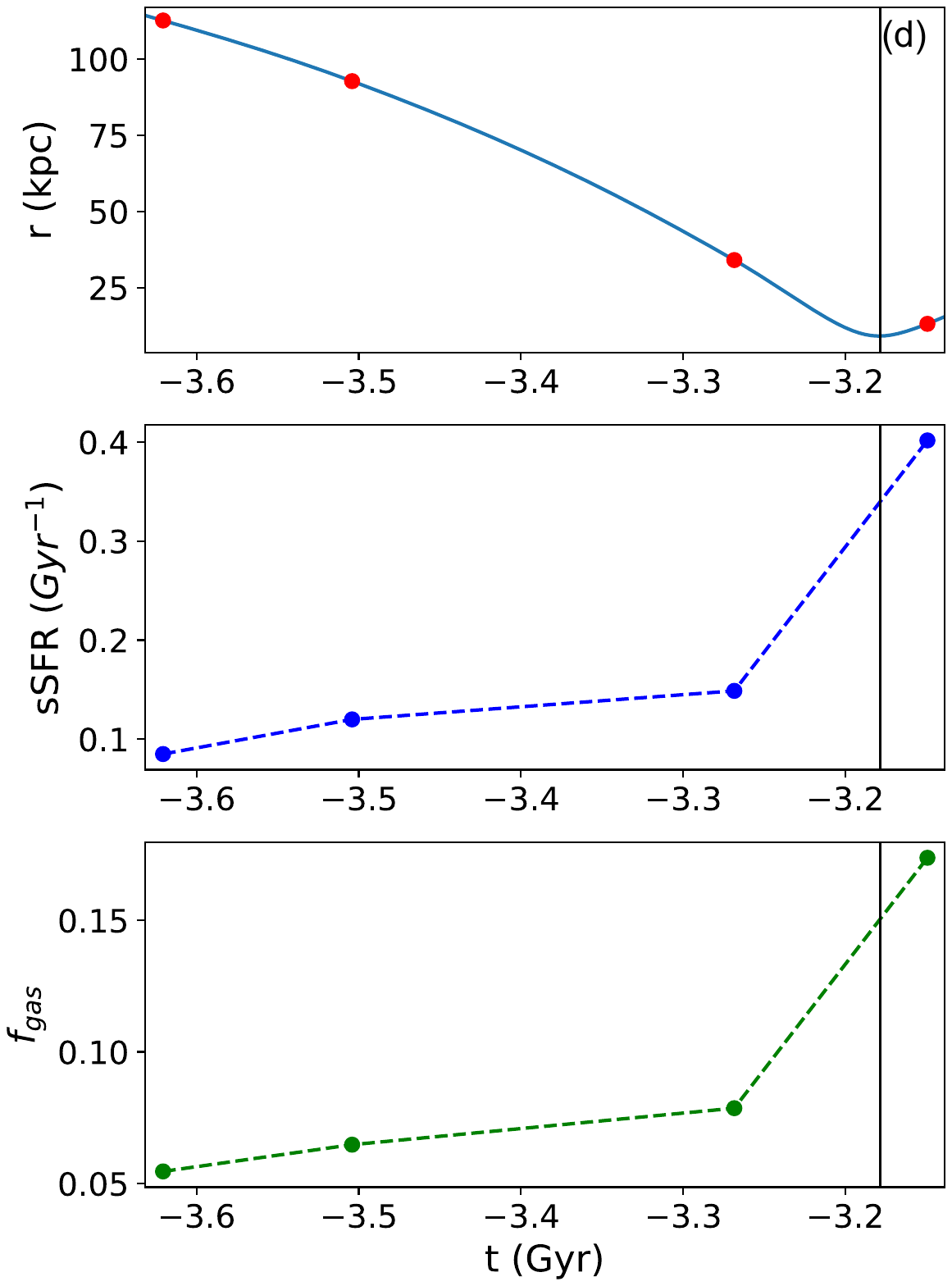}
  \end{minipage}

  \caption[Four examples of encounters]{Four examples of encounters in our final dataset. On the top panel of each plot is the snapshot data (red circles), as well as the 6D kinematic interpolation (blue line). On the second and third panels of each plot, the $sSFR$ ($\rm Gyr^{-1}$) and $\rm f_{gas}$ within $\rm R_{1/2}$ for the host galaxy are plotted (blue and green circles, respectively), with the dashed lines added for readability and not representing any quantitative interpolation between points. The vertical black solid line in each panel represents the time of pericentre.}
  \label{fig:encountexample}
\end{figure*}

In Figure \ref{fig:encountexample}, we examine how properties of the host galaxy, such as sSFR and $\rm f_{gas}$, change over the course of four representative encounters.
Encounters (a) and (b) show a medium-range infall, where the galaxy pairs start reasonably well-separated and undergo a pericentre at approximately half that initial separation.
Encounter (c) presents an extended encounter, where the galaxy pair goes from a close pericentre to a much larger apocentre. 
Encounter (d) shows an interacting galaxy pair in a shrinking encounter, with only one post-pericentre data point, likely due to another close pericentre or merger occurring within two snapshots after the pericentre considered.

In each encounter depicted in Figure \ref{fig:encountexample}, we find that sSFR is enhanced following the pericentre.
Using high-resolution merger simulations, \citet{Torrey2012} and \citet{Moreno2015} find a significant SFR enhancement within 0.5 Gyr of the pericentre.
The degree of these sSFR enhancements in Figure \ref{fig:encountexample} varies per encounter, though we would hesitate to make any conclusions linking the specifics of these enhancements to the orbital parameters of the individual encounters for such a small sample. 
In each encounter, $\rm f_{gas}$ is also enhanced alongside sSFR. 
In encounters (c) and (d), the changes in sSFR and $\rm f_{gas}$ appear similar in relative size and timing.
This is in contrast to encounters (a) and (b), where, although sSFR tends to be higher when $\rm f_{gas}$ is higher, they do not exhibit a consistent variation.
This suggests $\rm f_{gas}$ may not be the only factor responsible for driving the changes in star formation activities for interacting galaxies.

\subsection{Encounter-Triggered sSFR Enhancements}\label{subsec:sSFR}
Moving beyond the individual encounters in Figure \ref{fig:encountexample}, we can calculate and plot the average value of sSFR relative to the pericentre times for all host galaxies in the encounters dataset. 
We categorise our galaxies into $r_{\text{peri}}$ bins as follows: 0-50 kpc for the closest encounters, 50-100 kpc for intermediately separated encounters, 100-200 kpc to include encounters at the largest distances where galaxy pairs still show enhancements \citep{Patton2013, Patton2020}, and 200-500 kpc for encounters with such large separations that interactions are unlikely to trigger changes in galaxy properties.

This statistical methodology offers numerous advantages compared to the analysis of reconstructed orbits.
By stacking our large dataset of encounters, we are able to detect changes in the properties of interacting galaxies over smaller timescales compared to the time intervals between the simulation snapshots.
Additionally, this approach allows us to mitigate the stochastic effects within individual interaction orbits (e.g. see Figure \ref{fig:encountexample}).
We examine the maximum enhancement in each property by comparing the average value throughout the encounter to the average initial value. 
This is quantified by: 
\begin{equation}\label{eq:Qmax}
   Q(sSFR) = \frac{sSFR}{sSFR_{\rm i}}, 
\end{equation}
where the "i" subscript denotes the average initial value of the property in the first $\Delta$t bin, where we expect the galaxies to be observed well before the encounter in question. 
Our definition of Q(sSFR) differs from that used in previous papers in this series, as it is defined relative to its own initial value rather than in comparison to a control sample.

\begin{figure}
    \centering
    \includegraphics[width=\columnwidth]{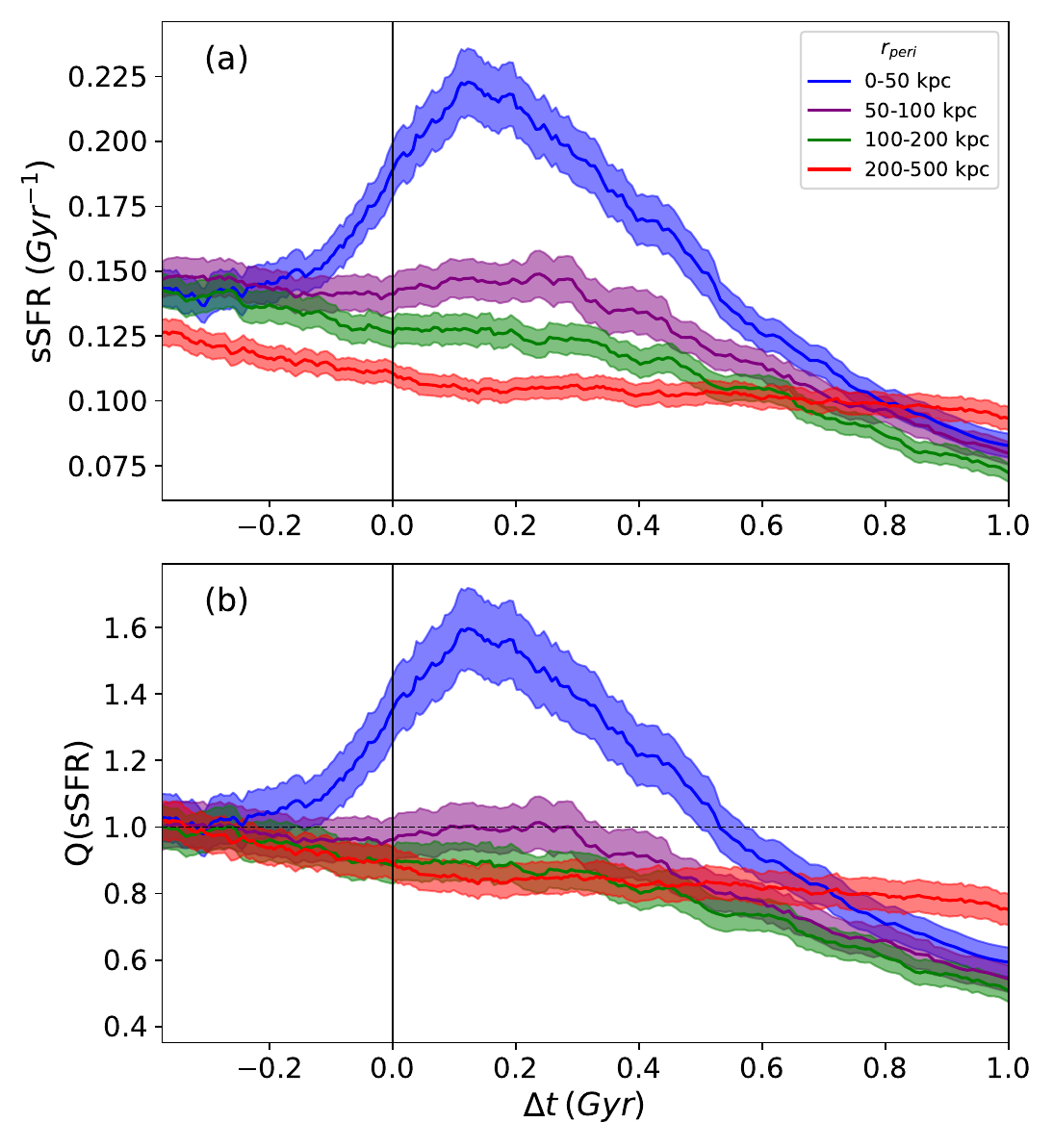}
    \caption{The upper panel shows the mean sSFR versus time relative to the pericentre ($\Delta t$) within R\textsubscript{1/2} for all host galaxies.
    The different colours represent various r\textsubscript{peri} bins. The black vertical line at $\Delta$t = 0 represents the time of the pericentre. The shaded regions represent the $2\sigma$ standard error in the mean. All averages have been smoothed using our fixed-width box kernel with a bin width of 3000 galaxies.
    The lower panel shows the enhancement, Q(sSFR) versus $\Delta t$.
    The black horizontal line at Q(sSFR) = 1 denotes the normalised mean pre-encounter value at the first snapshot.}
    \label{fig:sSFRHalf}
\end{figure}

Figure \ref{fig:sSFRHalf} presents the average sSFR against the time relative to the pericentre ($\Delta$t) (upper panel) and the relative sSFR enhancement (Q(sSFR); lower panel), within the stellar half-mass radius (R\textsubscript{1/2}) for simulated TNG100-1 host galaxies.
The stellar half-mass radius, R\textsubscript{1/2}, is the radius containing half of the total stellar mass associated with a galaxy.
Galaxies within our dataset have an average R\textsubscript{1/2} of 4 kpc.
Observations \citep{Ellison2013, BarreraBallesteros2015, Pan2019, Thorp2019, Thorp2024} and merger simulations \citep{DiMatteo2007, Moreno2015, Moreno2021} indicate that star formation enhancements due to encounters tend to be centrally concentrated.
As such, we expect to capture a majority of any enhanced star formation within R\textsubscript{1/2}.

In the lower panel of Figure \ref{fig:sSFRHalf}, we find a distinct enhancement in sSFR following the encounter in the 0-50 kpc bin.
For these closest encounters, mean sSFR rises as the galaxies approach pericentre before reaching a maximum value of $1.6 \pm 0.1$ times its first pre-encounter value at approximately 0.12 Gyr after the pericentre. 

\begin{figure}
    \centering
    \includegraphics[width=\columnwidth]{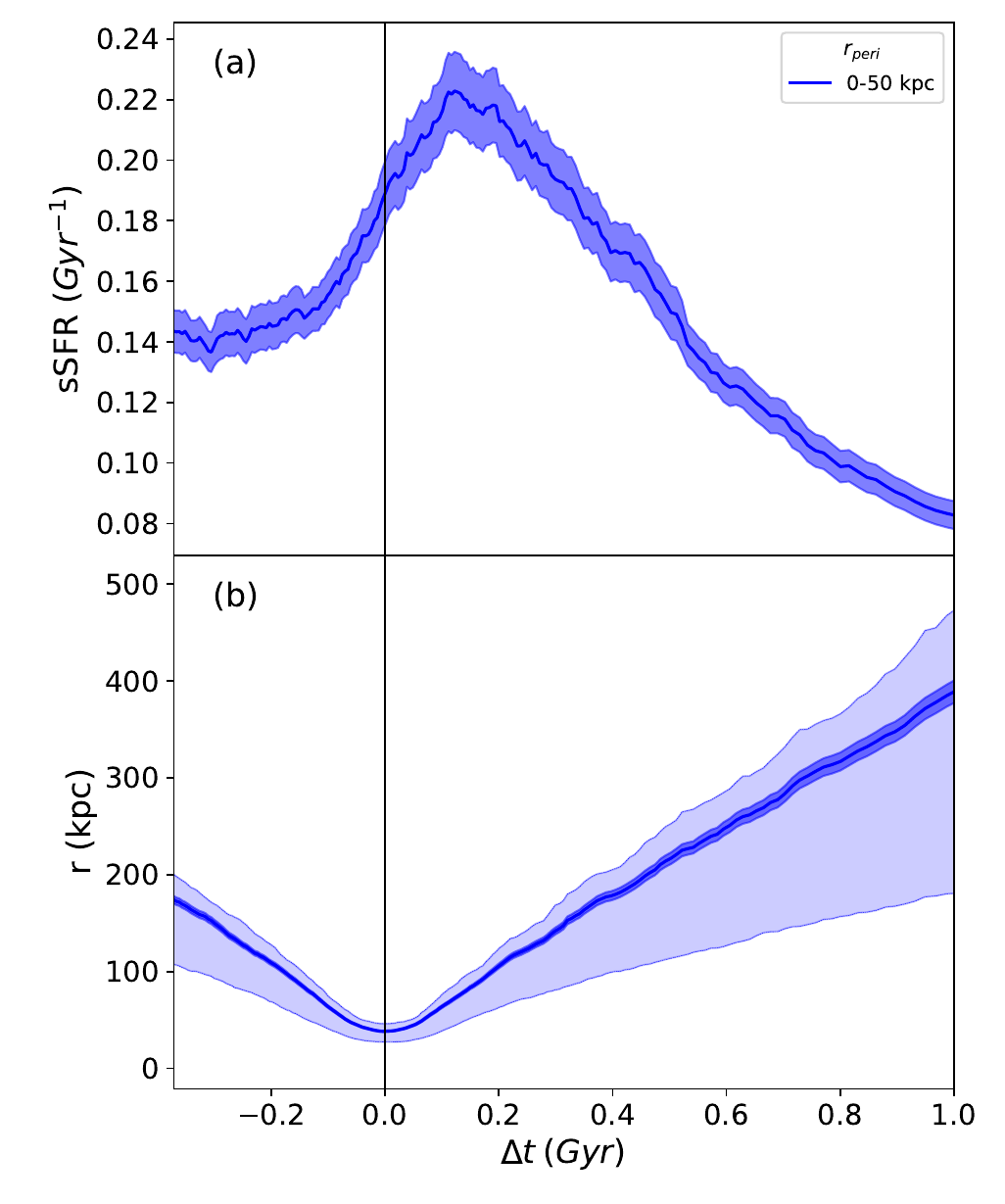}
    \caption[Comparing mean sSFR separation for host galaxies with r\textsubscript{peri} = 0-50 kpc]{Comparing average sSFR and separation for host galaxies with r\textsubscript{peri} = 0-50 kpc. The upper panel shows the mean sSFR versus time relative to pericentre within R\textsubscript{1/2}. The lower panel shows the mean pair separation bounded by the 25th and 75th percentiles. The black vertical line at $\Delta$t = 0 represents the time of pericentre.}
    \label{fig:sSFRrpericompare}
\end{figure}

In Figure \ref{fig:sSFRrpericompare}, we plot the mean sSFR in the 0-50 kpc bin across the pericentre from Figure \ref{fig:sSFRHalf}, as well as the mean separation of these galaxy pairs.
We note that on average in the 0-50 kpc r\textsubscript{peri} bin, galaxies have a mean initial separation of $\sim$175 kpc and a mean separation $\sim$30 kpc at pericentre. 
At the time of maximum sSFR enhancement at $\Delta$t = 0.12 Gyr, the galaxies have a mean separation of $\sim$75 kpc.
The sSFR in this r\textsubscript{peri} bin then sharply drops off, decreasing below its pre-encounter value at $\sim$0.52 Gyr, eventually reaching an average value lower than the passively evolving 200-500 kpc bin.
This is possibly due to the enhanced star formation rate triggered by the pericentric interactions suppressing the star formation of the galaxy relative to a passively evolving population through exhaustion of the gas supply, star formation feedback, or other quenching mechanisms.

We find a correlation between sSFR and r\textsubscript{peri}, whereby galaxies involved in encounters with a smaller r\textsubscript{peri} have a higher post-encounter sSFR (upper panel of Figure \ref{fig:sSFRHalf}) and a higher sSFR enhancement (lower panel of Figure \ref{fig:sSFRHalf}). 
Within these r\textsubscript{peri} bins, the 200-500 kpc bin of encounters is so far separated at their pericentric passage that they display passive evolution. 
While sSFR in the 50-100 kpc and 100-200 kpc bins is generally higher than the 200-500 kpc bin, it is difficult to concretely identify an enhancement given the uncertainty.

\begin{figure}
    \centering
    \includegraphics[width=\columnwidth]{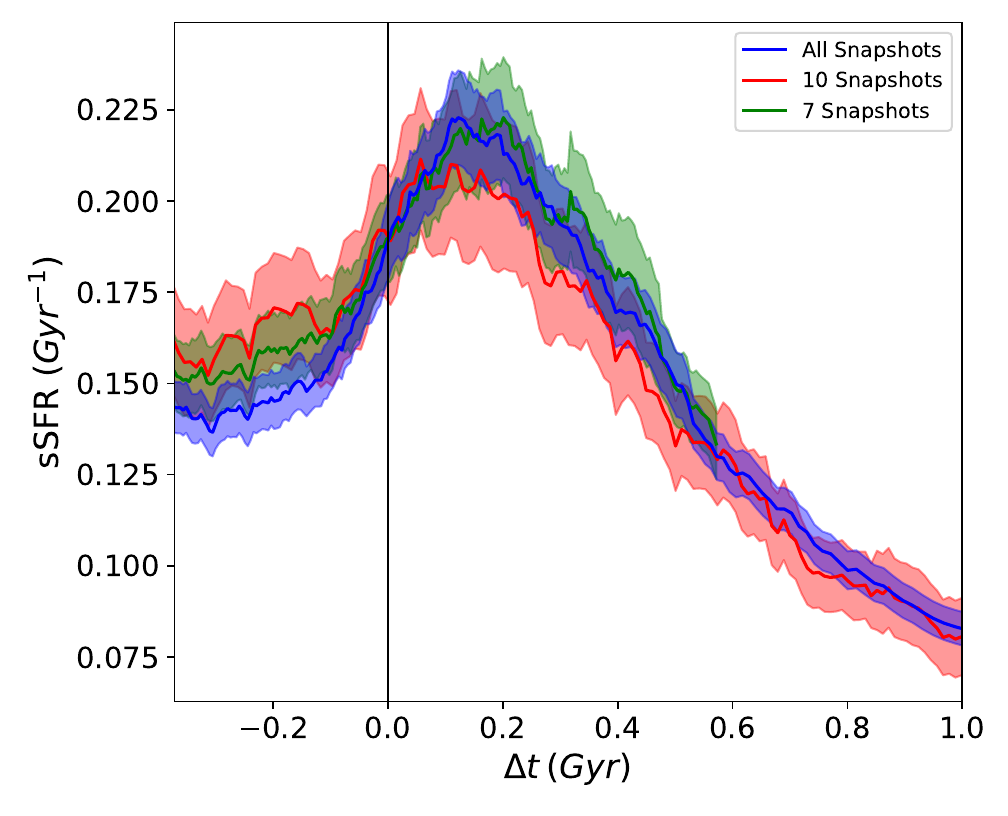}
    \caption[Comparing mean sSFR separation for host galaxies with r\textsubscript{peri} = 0-50 kpc]{Mean sSFR within R\textsubscript{1/2} for host galaxies with r\textsubscript{peri} = 0-50 kpc (blue) compared to mean sSFR when limited to the first 7 snapshots of all encounters with at least 7 snapshots (green), and the first 10 snapshots of all encounters with at least 10 snapshots (red).}
    \label{fig:dropout}
\end{figure}

Due to our selection criteria as described in Section \ref{subsec:encount}, encounters within our dataset have differing numbers of snapshots. 
As such, as we move right along the $\Delta t$ axis in Figure \ref{fig:sSFRHalf}, shorter-lived encounters drop out of the moving average.
It is possible that these dropouts may affect the averages calculated if they introduce a bias towards longer encounters, which might not be representative of the entire population. 
Since most encounters within our dataset have 4-10 snapshots, understanding the impact of these dropouts is crucial to ensure that our results are not skewed by the uneven temporal coverage of encounters.

In Figure \ref{fig:dropout}, we compare the average sSFRs taken with 7 and 10 snapshots with the full average for $r_{\text{peri}} = 0-50$ kpc. 
Each snapshot-limited average is complete across the entire $\Delta t$ range; i.e., unlike the full sample, there are no dropouts, and there are the same number of galaxies in each bin. This allows us to directly assess the impact of dropouts on our calculations.
The snapshot-limited samples more than adequately cover the 0.2 Gyr period following the pericentre, where we find our maximum enhancements.
We do not find any concerning deviations in the averages of the snapshot-limited samples from the average of the full sample, suggesting that the dropout effect does not significantly impact our results. This consistency indicates that our analysis is robust and that the sSFR trends we observe are reliable, even with varying numbers of snapshots across encounters.

\subsection{Enhancement of \texorpdfstring{f\textsubscript{gas}}{fgas} and \texorpdfstring{SFE\textsubscript{H}}{SFEH}}\label{subsec:galprop}
\begin{figure*}
    \centering
    \includegraphics[width=\textwidth]{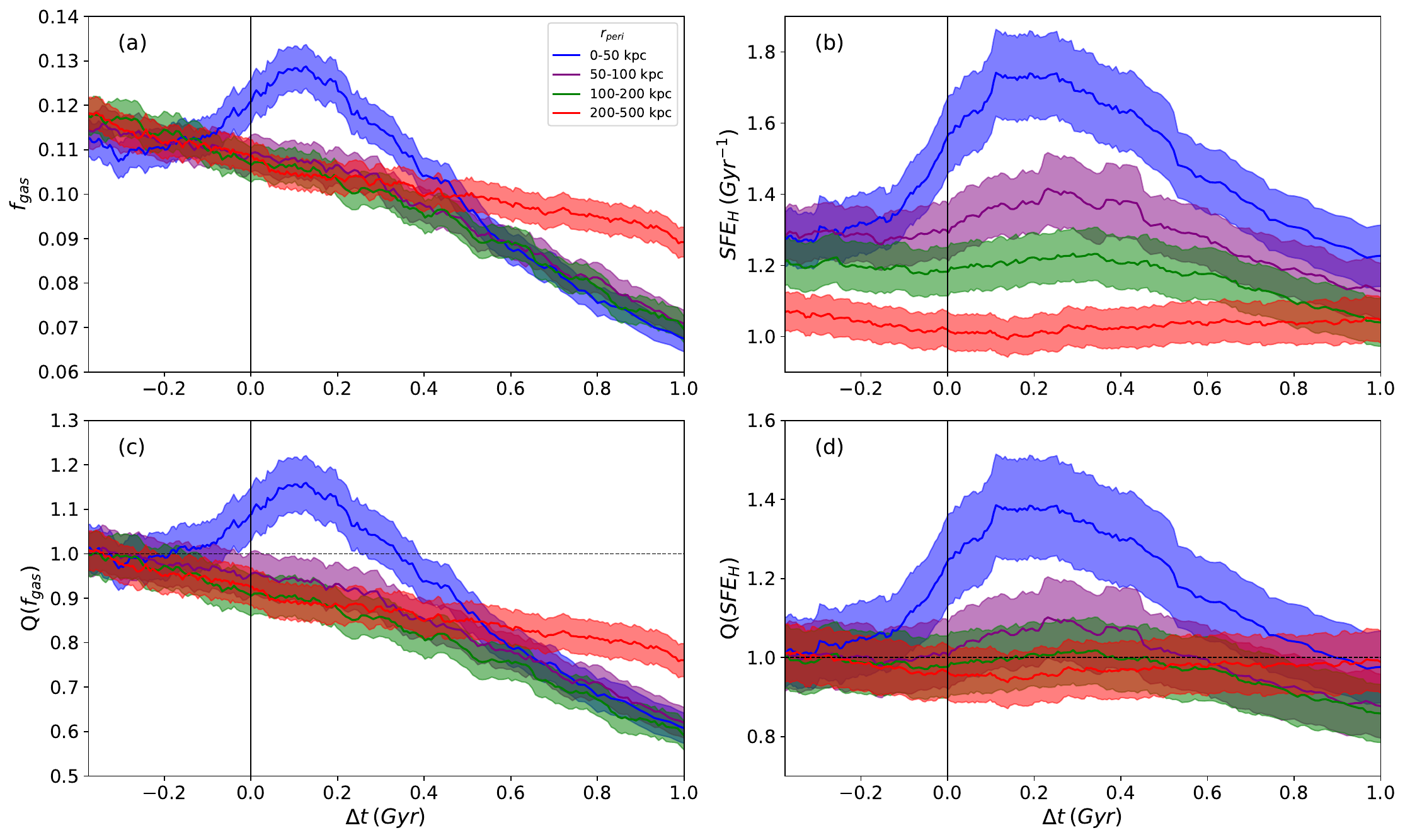}
    \caption{Mean galaxy properties of interest versus time relative to the pericentre within R\textsubscript{1/2} for all host galaxies. 
    The second row shows the enhancements (Q) of these properties versus time relative to the pericentre.
    The different colours represent various r\textsubscript{peri} bins. The black vertical line at $\Delta$t = 0 represents the time of the pericentre. The shaded regions represent the $2\sigma$ standard error in the mean. All averages have been smoothed using our fixed-width box kernel with a bin width of 3000 galaxies. The shaded regions represent the $2\sigma$ standard error in the mean.}
    \label{fig:propshalf1}
\end{figure*}

In Section \ref{subsec:sSFR}, we found that sSFR is clearly enhanced in close encounters (0-50 kpc).
To better understand the drivers for encounter-triggered star formation, we examine $\rm f_{gas}$ and $\rm SFE_{H}$. 
Figure \ref{fig:propshalf1} presents the average $\rm f_{gas}$ (panel (a)) and $\rm SFE_{H}$ (panel (b)), respectively, against the time relative to the pericentre ($\Delta$t) within the stellar half-mass radius (R\textsubscript{1/2}) for simulated TNG100-1 host galaxies. 
Focusing on measurements within $R_{1/2}$ helps to minimise contributions from hot gas to $\rm f_{gas}$, ensuring this metric better traces gas phases relevant to star formation \citep{Diemer2019}.
We calculate $\rm SFE_{H}$ using the average sSFR and $\rm f_{gas}$ according to Equation \ref{eq:fineq}. 
While this approach introduces greater uncertainty in $\rm SFE_{H}$ compared to calculating it directly—due to the propagation of uncertainties from the averaged quantities—it effectively mitigates potential singularities in galaxies where $\rm f_{gas} = 0$.

In panel (a) and (c) of Figure \ref{fig:propshalf1}, we see a distinct enhancement of $\rm f_{gas}$ within the stellar half-mass radius in the 0-50 kpc r\textsubscript{peri} bin. 
This suggests an inflow of gas towards the central regions due to encounters. 
This enhancement reaches a maximum value of $1.2 \pm 0.1$ times its first pre-encounter value at $\sim$0.12 Gyr, as shown in panel (c), which coincides with the time of maximum sSFR enhancement. 
The $\rm f_{gas}$ returns to its pre-encounter value at $\Delta t =$ 0.36 Gyr, resulting in a shorter-lived enhancement than that observed in sSFR in panel (b) of Figure \ref{fig:sSFRHalf}.
The remaining r\textsubscript{peri} 50-100 kpc and 100-200 kpc bins display no enhancement in $\rm f_{gas}$, and instead follow the passive evolution of the 200-500 kpc r\textsubscript{peri} bin until approximately 0.2 Gyr, where they are then suppressed below the 200-500 kpc bin. 
This is consistent with the higher sSFR in these r\textsubscript{peri} bins, as seen in panel (a) of Figure \ref{fig:sSFRHalf}, potentially expediting the depletion of the gas reservoir in these galaxies. 

Panels (b) and (d) of Figure \ref{fig:propshalf1} show a distinct enhancement of SFE\textsubscript{H} in the 0-50 kpc r\textsubscript{peri} bin, reaching a maximum value of $1.4 \pm 0.1$ times the pre-encounter value, which it roughly maintains between $\Delta t = 0.1$ Gyr and $\Delta t = 0.2$ Gyr before decreasing to its pre-encounter value 0.8 Gyr after the pericentre. 
This shows that enhancements of $\rm SFE_{H}$ due to the pericentre are significantly more long-lived than enhancements in gas supply. 
We observe that higher average $\rm SFE_{H}$ is correlated with lower r\textsubscript{peri}, as evidenced by the ordered separation of r\textsubscript{peri} bins.
The higher $\rm SFE_{H}$ in r\textsubscript{peri} bins larger than 50 kpc may help to explain the higher sSFR in these bins, seen in Figure \ref{fig:sSFRHalf}, despite the apparent lack of enhancement in gas supply seen in panel (c) of Figure \ref{fig:propshalf1}. 
Alternatively, should a galaxy pair be engaged in a prolonged multi-pericentre shrinking orbit, it is also feasible that previous encounters at wider r\textsubscript{peri} could result in overall higher sSFR for later encounters at closer r\textsubscript{peri}.

sSFR, $\rm f_{gas}$ and $\rm SFE_{H}$ are intimately connected properties, as shown in Equation \ref{eq:fineq}.
As such, by examining how they change, we can determine the relative importance of $\rm f_{gas}$ and $\rm SFE_{H}$ in driving changes in sSFR.
From Figures \ref{fig:sSFRHalf} and \ref{fig:propshalf1}, we find that sSFR, $\rm f_{gas}$ and $\rm SFE_{H}$ are all significantly enhanced following the pericentre of a galaxy-galaxy encounter in the 0-50 kpc r\textsubscript{peri}. 
In all three properties, these enhancements reach a maximum at approximately 0.1 Gyr following the pericentre. 
We find that relative to the $\rm f_{gas}$, the enhancement in $\rm SFE_{H}$ is larger and longer-lived. 
This suggests that $\rm SFE_{H}$ plays an outsized role in producing the observed enhancement in sSFR within $R_{1/2}$.

\begin{figure*}
    \centering
    \includegraphics[width=\textwidth]{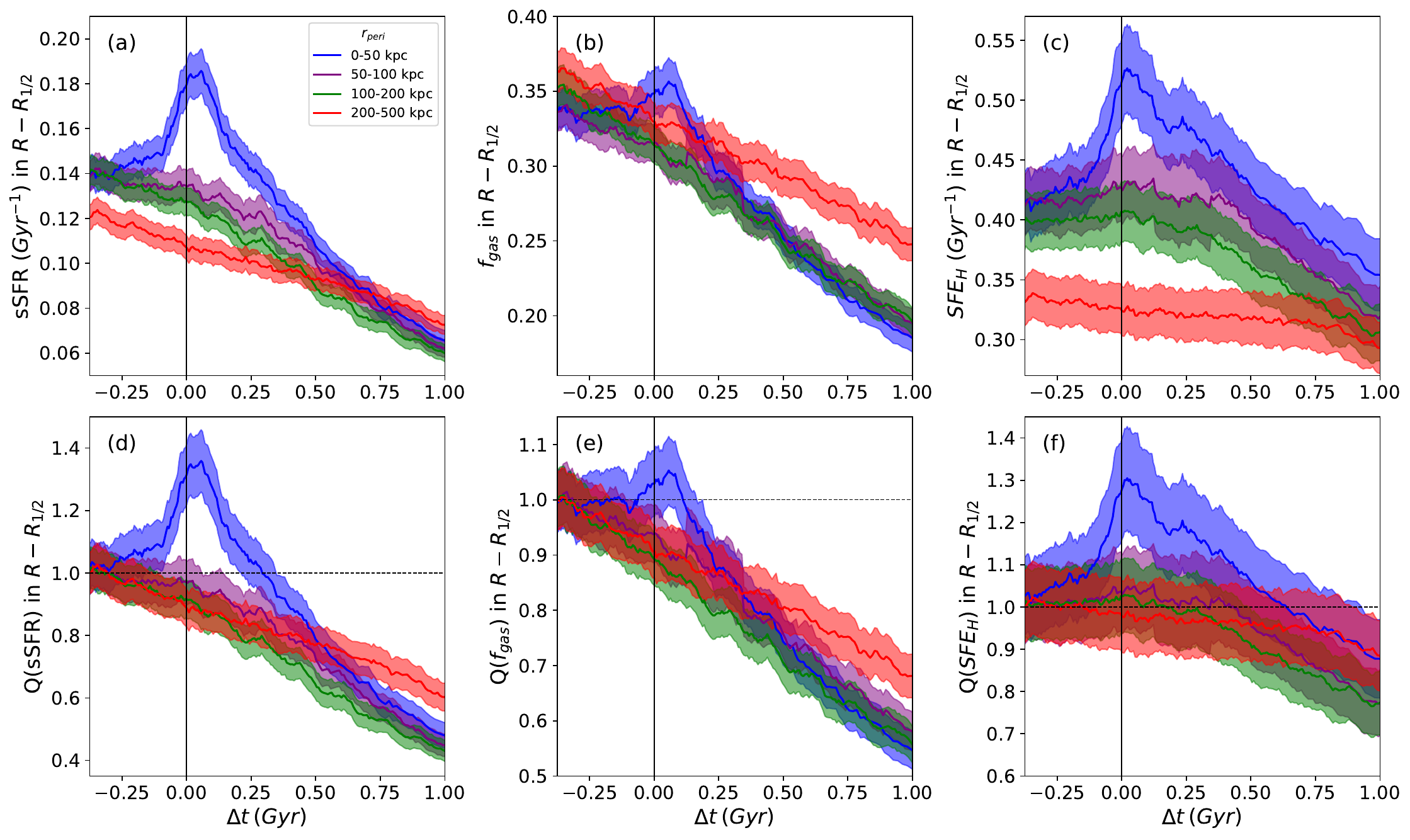}
    \caption{Mean galaxy properties of interest versus time relative to the pericentre within outer stellar half-mass radius shell (R - R\textsubscript{1/2}) for all host galaxies. 
    The second row shows the enhancements (Q) of these properties versus time relative to the pericentre.
    The different colours represent various r\textsubscript{peri} bins. The black vertical line at $\Delta$t = 0 represents the time of the pericentre. The shaded regions represent the $2\sigma$ standard error in the mean. All averages have been smoothed using our fixed-width box kernel with a bin width of 3000 galaxies. The shaded regions represent the $2\sigma$ standard error in the mean.}
    \label{fig:propsouter}
\end{figure*}

Figure \ref{fig:propsouter} presents the average sSFR (panel (a)), $\rm f_{gas}$ (panel (b)) and $\rm SFE_{H}$ (panel (c)), respectively, plotted against $\Delta$t, within the outer shell (R - R\textsubscript{1/2}) of host galaxies.
This shell is defined by the remaining volume when the volume enclosed by the stellar half-mass radius (R\textsubscript{1/2}) is subtracted from the volume enclosed by twice the stellar half-mass radius (R) and is meant to represent the outer region of the galaxy. 
This outer shell will typically enclose 20-30 per cent of the galaxy's stellar mass \citep{Genel2018}.

We examine the evolution of the sSFR, $\rm f_{gas}$ and $\rm SFE_{H}$ in this region to better spatially contextualise the effects of the encounter on these properties. 
Here we observe similar features to those discussed earlier for Figure \ref{fig:propshalf1}.
There is an enhancement in all three properties following the pericentre. 
However, it is worth noting that, on average, we observe lower enhancements across all properties in comparison to the inner stellar half-mass radius.
Additionally, the maximum enhancements in the 0-50 kpc r\textsubscript{peri} bins occur significantly earlier following the pericentre than in Figure \ref{fig:propshalf1}.
This suggests a time lag in the response of the galactic properties to the pericentre from the outskirts to the inner region of the galaxy, whereby the sSFR, $\rm f_{gas}$ and $\rm SFE_{H}$ reach a maximum approximately 0.05 Gyr earlier than within the central stellar half-mass radius. 
This may be indicative of inflows of gas from the outer region of the galaxy toward the centre. 
These inflows can bring a fresh supply of gas to the galactic centre while the movement of the gas may enhance the efficiency at which the gas is turned into stars.

A comparison of the upper panels of Figures \ref{fig:sSFRHalf} and \ref{fig:propshalf1} to panels (a), (b), and (c) of Figure \ref{fig:propsouter} suggests a scenario where close encounters trigger an enhancement in the galaxy's properties which begin in the outer disk and then travels inward, centrally concentrating within the stellar half-mass radius. 
These results suggest that gas from the outer regions of the galaxy flows into the region defined by the outer shell, triggering a small enhancement in sSFR, $\rm f_{gas}$ and $\rm SFE_{H}$.
This gas continues to flow inward, concentrating within $R_{1/2}$, suppressing star formation in the outer shell and triggering relatively larger enhancements in the central region of the galaxy. 

In Figure \ref{fig:sSFRHalf} for the inner stellar half-mass radius as well as Figure \ref{fig:propsouter} for the outer shell, we note a systematic decline in the sSFR, $\rm f_{gas}$ and $\rm SFE_{H}$ at large $\Delta$t. 
This is most likely explained by passive evolution of galaxies engaged in long-lived encounters.

\subsection{Metallicity}\label{sec:metallicty}

\begin{figure*}
     \centering
     \begin{subfigure}{0.49\textwidth}
         \centering
         \includegraphics[width=\textwidth]{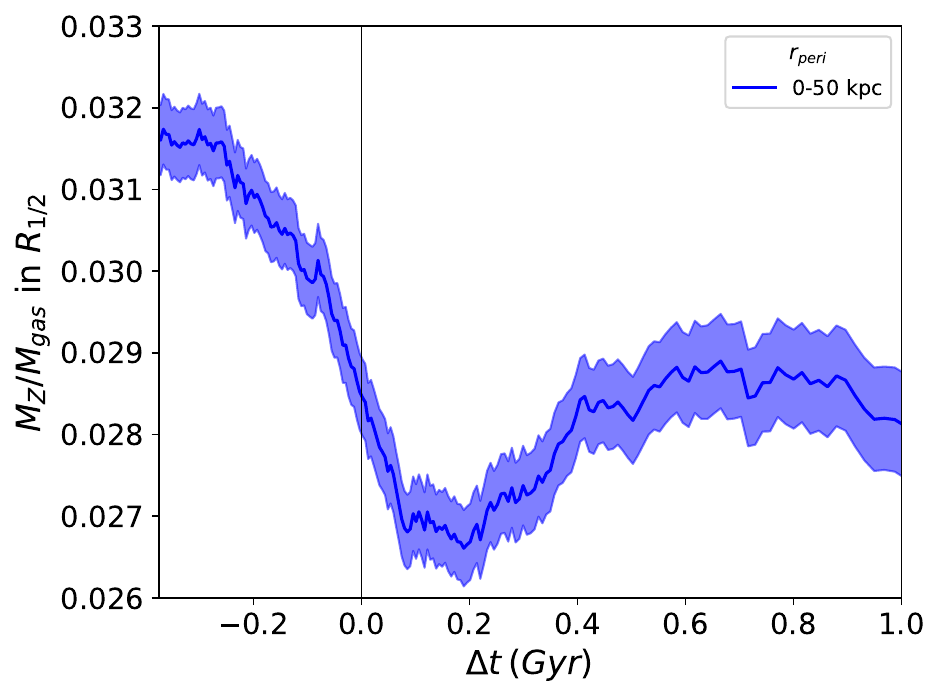}
         \caption{$\rm M_{Z}/M_{gas}$ in $\rm R_{1/2}$}
         \label{fig:metalhalf}
     \end{subfigure}
     \hfill
     \begin{subfigure}{0.49\textwidth}
         \centering
         \includegraphics[width=\textwidth]{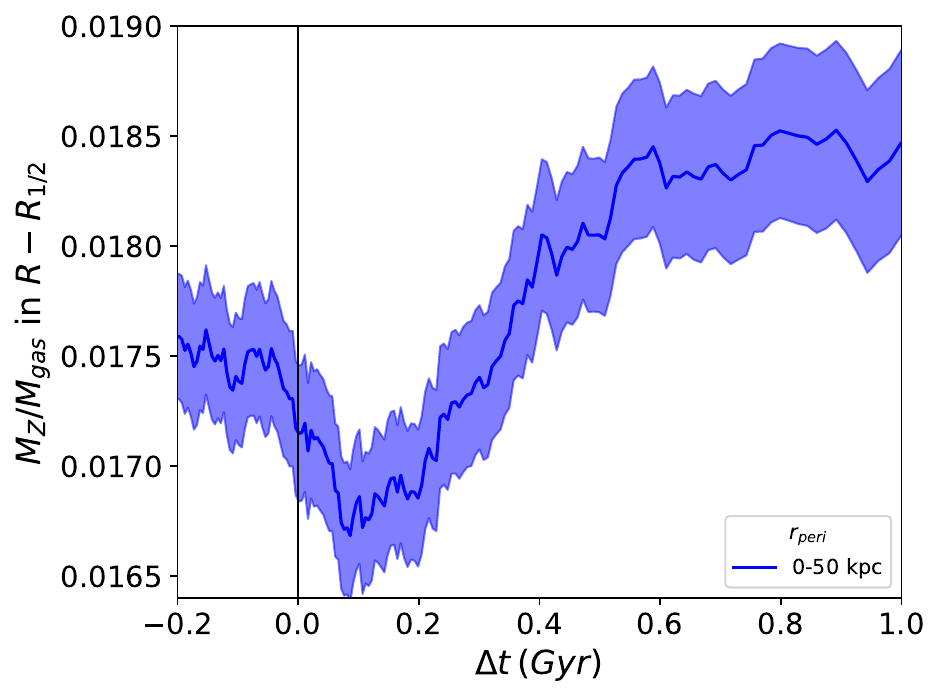}
         \caption{$\rm M_{Z}/M_{gas}$ in $\rm R-R_{1/2}$}
         \label{fig:metalshell}
     \end{subfigure}
        \caption{Mean gas metallicity versus time relative to pericentre within $\rm R_{1/2}$ (left panel) and R - R\textsubscript{1/2} (right panel) for host galaxies with r\textsubscript{peri} = 0-50 kpc. The black vertical line at $\Delta$t = 0 represents the time of the pericentre. The shaded regions represent the $2\sigma$ standard error in the mean.}
        \label{fig:metal}
\end{figure*}

To better understand the nature of the inflowing gas indicated in Section \ref{subsec:galprop}, we examine the metallicity of the gas.
This gas metallicity, $\rm M_{Z}/M_{gas}$, is defined for the inner stellar half-mass radius and the outer shell as the unit-less ratio of the mass of metals to total gas mass within those regions. 

In Figure \ref{fig:metal} we plot $\rm M_{Z}/M_{gas}$ over the course of the encounter for the 0-50 kpc r\textsubscript{peri} bin of (a) the inner stellar half-mass radius and (b) the outer shell of host galaxies. 

In panel (a) of Figure \ref{fig:metal}, we observe that the gas metallicity drops precipitously from a pre-encounter value of 0.0315 to a minimum of 0.027 at around 0.1 - 0.2 Gyr. This minimum in gas metallicity aligns with the times of maximum enhancement in sSFR, $\rm f_{gas}$ and $\rm SFE_{H}$ seen in Figures \ref{fig:sSFRHalf} and \ref{fig:propshalf1}. 
An inflow of pristine star-forming gas toward the centre of the galaxy is a likely explanation for this sudden drop in metallicity \citep{Rupke2010, BarreraBallesteros2015, Bustamante2020, Sparre2022}. 
Beyond $\Delta t = 0.2$ Gyr, the gas metallicity increases marginally, likely due to star formation enrichment, but is still diluted compared to its pre-encounter value. 
This dilution persists even as enhancements in sSFR and $\rm f_{gas}$ wane. 

In panel (b) of Figure \ref{fig:metal} for the outer shell, we observe a similar, but much smaller, drop in the metallicity following the encounter. 
However, following the minimum in metallicity at approximately 0.1 Gyr, the metallicity increases past its pre-encounter value to a maximum of 0.0185, likely due to star formation enrichment.
It is noteworthy that the average gas metallicity in the outer shell is significantly lower than in the inner stellar half-mass radius.
This lower average gas metallicity in the outer shell is consistent with \citet{Torrey2019}, who found a metallicity gradient where the gas metallicity decreases with radius for IllustrisTNG galaxies.
But the difference in average metallicity does suggest that inflows from this shell could be capable of explaining some of the dilution in the inner stellar half-mass radius.
However, since there is also evidence of dilution in the outer shell, some inflows of pristine gas must have come from beyond twice the stellar half-mass radius of the galaxy. 

\begin{figure}
    \includegraphics[width=0.5\textwidth]{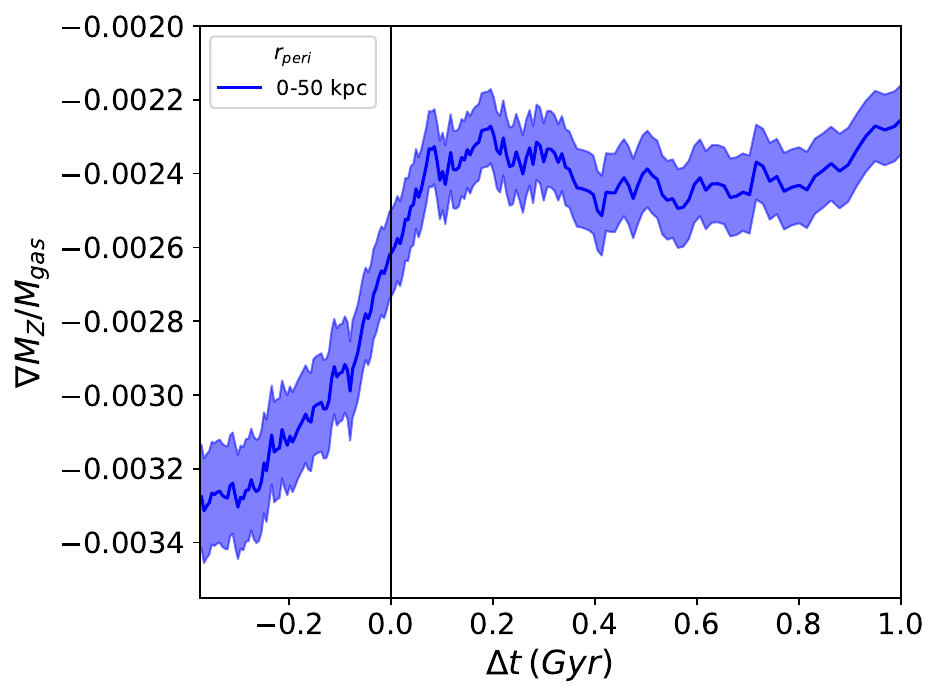}
    \caption[Metallicity Gradient]{Mean gas metallicity gradient versus time relative to pericentre calculated using the metallicity in $\rm R_{1/2}$ (Figure \ref{fig:metalhalf}) and R - R\textsubscript{1/2} (Figure \ref{fig:metalshell}) for host galaxies with r\textsubscript{peri} = 0-50 kpc. The black vertical line at $\Delta$t = 0 represents the time of the pericentre. The shaded regions represent the $2\sigma$ standard error in the mean.}
    \label{fig:metalgrad}
\end{figure}

In Figure \ref{fig:metalgrad}, we plot a crude metallicity gradient using the enclosed metallicity within in $\rm R_{1/2}$ and R - R\textsubscript{1/2} from Figure \ref{fig:metal} and the average value of $\rm R_{1/2} \approx 4 kpc$ for host galaxies with r\textsubscript{peri} $\leq$ 50 kpc.
The metallicity gradient, initially negative prior to pericentre, reflects the typical radial structure of galaxy metallicity, with higher enrichment at smaller radii \citep{Ho2015, Belfiore2017, Carton2018}. 
However, approaching pericentre ($\Delta t = 0$), the gradient increases sharply, as the inner region experiences the significant dilution due to inflowing metal-poor gas shown in Figure \ref{fig:metalhalf}, while the outer region sees a comparably minor dilution. 
The peak of this gradient again aligns with the times of maximum enhancement in sSFR, $\rm f_{gas}$ and $\rm SFE_{H}$ seen in Figures \ref{fig:sSFRHalf} and \ref{fig:propshalf1} after which the gradient roughly flattens out.
From Figures \ref{fig:metal} and \ref{fig:metalgrad}, we can estimate a rough metallicity depletion timescale of 0.4 Gyr between $\rm \Delta t = -0.2$, where the metallicity begins to drop, and $\rm \Delta t = 0.2$, after which it flattens out.

\subsection{Fuelled Fraction of Star Formation}\label{subsec:fuelfrac}
We find a clear increase in sSFR during close encounters. 
Since sSFR can be expressed as a product of $\rm f_{gas}$ and $\rm SFE_{H}$, we aim to quantify the relative contribution of an enhancement in $\rm f_{gas}$ and in $\rm SFE_{H}$ to the observed enhancement in sSFR.
In doing so, we hope to determine whether fuelling or efficiency drives enhanced star formation in interacting galaxies.

The specific quantity we aim to split into a fuelled or efficiency-driven component is the fractional change in sSFR at time $\Delta t$. 
This is equal to the change in sSFR (hereafter $\delta sSFR = sSFR - sSFR_{i}$) divided by the initial sSFR (sSFR\textsubscript{i}) as follows:

\begin{equation}\label{eq:QsSFRnormdef}
    \frac{\delta sSFR}{sSFR_{i}} = \frac{sSFR-sSFR_{i}}{sSFR_{i}} = Q(sSFR) - 1
\end{equation}

We can then split the relative change in sSFR ($\rm \delta sSFR/sSFR_{i}$) into two terms that depend separately on the relative change in f\textsubscript{gas} and SFE\textsubscript{H}.
From Equation \ref{eq:fineq}, we can define:

\begin{equation}
    Q(sSFR) = Q(f_{gas}) \cdot Q(SFE_{H})
\end{equation}
Taking the log of both sides:
\begin{equation}
    \log Q(sSFR) = \log Q(f_{gas}) + \log Q(SFE_{H})
\end{equation}
Normalising by log Q(sSFR):
\begin{equation}\label{eq:fuelfracpercent}
    \frac{\log Q(f_{gas})}{\log Q(sSFR)} + \frac{\log Q(SFE_{H})}{\log Q(sSFR)} = 1
\end{equation}

Equation \ref{eq:fuelfracpercent} yields two terms which sum to one and represent the fraction of $\rm \delta sSFR/sSFR_{i}$ which can be attributed to an enhancement in $\rm f_{gas}$ and $\rm SFE_{H}$, respectively.

From Equations \ref{eq:fuelfracpercent} and \ref{eq:QsSFRnormdef}, we can define the fuelled fraction of $\delta \rm sSFR/\rm sSFR_{\rm i}$, the amount of new star formation which can be attributed to the enhancement in $\rm f_{gas}$, as:

\begin{equation}\label{eq:fuelledfrac}
    \rm Fuelled~\frac{\delta sSFR}{sSFR_{\rm i}} = \frac{\log Q(\rm f_{gas})}{\log Q(sSFR)} \cdot \frac{\delta sSFR}{sSFR_{\rm i}}
\end{equation}
Consequently, the efficiency fraction, the amount of new star formation which can be attributed to the enhancement in $\rm SFE_{H}$, is:
\begin{equation}\label{eq:efficiencyfrac}
    \rm Efficiency~\frac{\delta sSFR}{sSFR_{\rm i}} = \frac{\delta sSFR}{sSFR_{\rm i}} - \rm Fuelled~\frac{\delta sSFR}{sSFR_{\rm i}}.
\end{equation}

\begin{figure*}
     \centering
     \begin{subfigure}{0.49\textwidth}
         \centering
         \includegraphics[width=\textwidth]{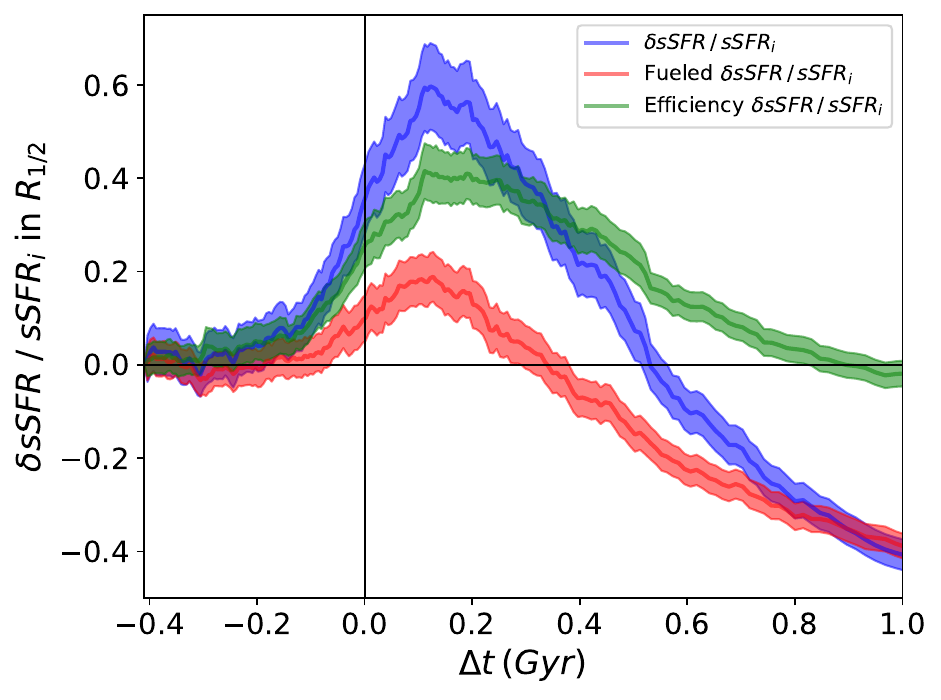}
         \caption{$\rm R_{1/2}$}
         \label{fig:fuelfrachalf}
     \end{subfigure}
     \hspace{0.01\textwidth} 
     \begin{subfigure}{0.49\textwidth}
         \centering
         \includegraphics[width=\textwidth]{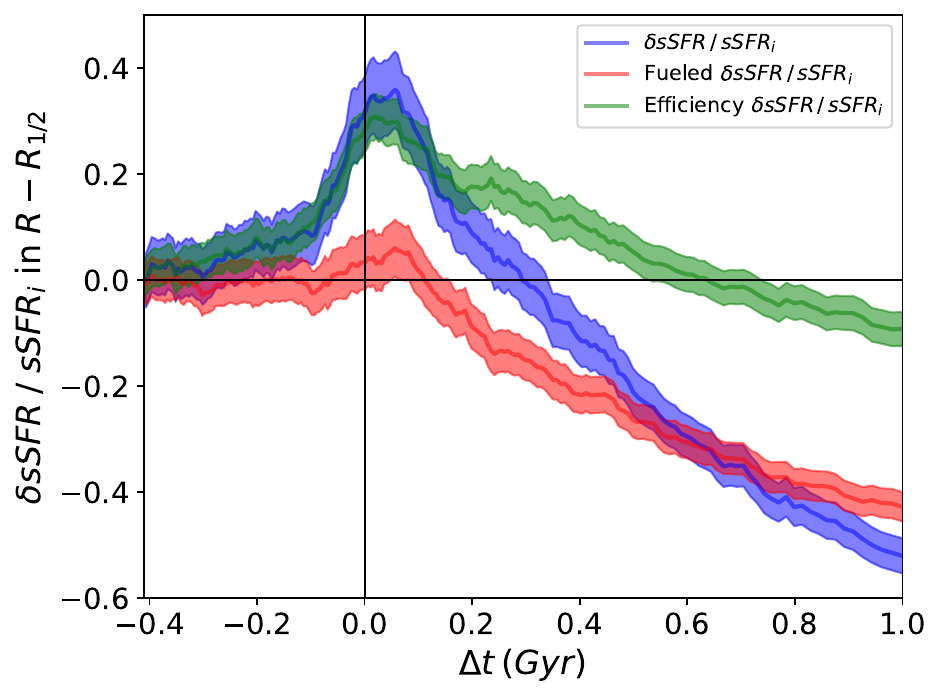}
         \caption{$\rm R-R_{1/2}$}
         \label{fig:fuelfracring}
     \end{subfigure}
     \caption{Enhancement in sSFR (blue), as well as the fraction driven by fuelling (red) and efficiency (green), versus time relative to pericentre within $R_{1/2}$ (left) and $R - R_{1/2}$ (right) for host galaxies with r\textsubscript{peri} = 0-50 kpc. The black vertical line at t\textsubscript{peri} = 0 represents the time at which the encounter occurs. The shaded regions represent the $2\sigma$ standard error in the mean computed using the Jackknife technique.}
     \label{fig:fuelfrac}
\end{figure*}

In Figure \ref{fig:fuelfrac}, we plot the enhancement in sSFR, normalised to an initial value of zero (represented by $\delta \rm sSFR / \rm sSFR_{\rm i}$), alongside the portion of that enhancement which may be attributed to enhanced fuelling (enhanced $\rm f_{gas}$, represented by fuelled $\delta \rm sSFR / \rm sSFR_{\rm i}$), and the portion which may be attributed to enhanced efficiency (enhanced $\rm SFE_{H}$, represented by efficiency $\delta \rm sSFR / \rm sSFR_{\rm i}$), for the 0-50 kpc r\textsubscript{peri} bin of inner stellar half-mass radius and the outer shell, respectively.
As in Sections \ref{subsec:sSFR} and \ref{subsec:galprop}, these are average properties calculated for the population described. 
In both the left and right panels of Figures \ref{fig:fuelfrac}, the initial enhancement in sSFR appears to be primarily driven by the increase in $\rm SFE_{H}$, with almost no change in $\rm f_{gas}$ during the earliest stages of the encounter (roughly 100-400 Myr before pericentre). 
This suggests that, before any augmentation of gas supply occurs, star formation becomes more efficient. 
As shown in the left panel of Figure \ref{fig:fuelfrac}, the contribution of fuel-driven star formation, represented by fuelled $\delta \rm sSFR / \rm sSFR_{\rm i}$, accounts for approximately 30 per cent of the peak enhancement in $\delta \rm sSFR / \rm sSFR_{\rm i}$. 
Additionally, the enhanced gas supply seems to be rapidly consumed, with the subsequent reduction in gas acting as a net drag on the system, leaving enhanced efficiency to sustain the elevated sSFR.

$\rm SFE_{H}$'s outsized role in driving star formation is even more pronounced in panel (b) of Figure \ref{fig:fuelfrac}, where the fuelled $\delta \rm sSFR / \rm sSFR_{\rm i}$ contributes less than 15 per cent of the maximum enhancement in $\delta \rm sSFR / \rm sSFR_{\rm i}$, giving the enhanced efficiency $\delta \rm sSFR / \rm sSFR_{\rm i}$ a more prominent role in counteracting the suppressive effects of gas depletion.

These results strongly suggest that the enhanced $\rm SFE_{H}$—representing either the increased efficiency with which star-forming gas is assembled, or the enhanced efficiency of star formation itself—acts as the primary driver for the elevated sSFR associated with close encounters.

\begin{figure}
    \centering
    \includegraphics[width=\columnwidth]{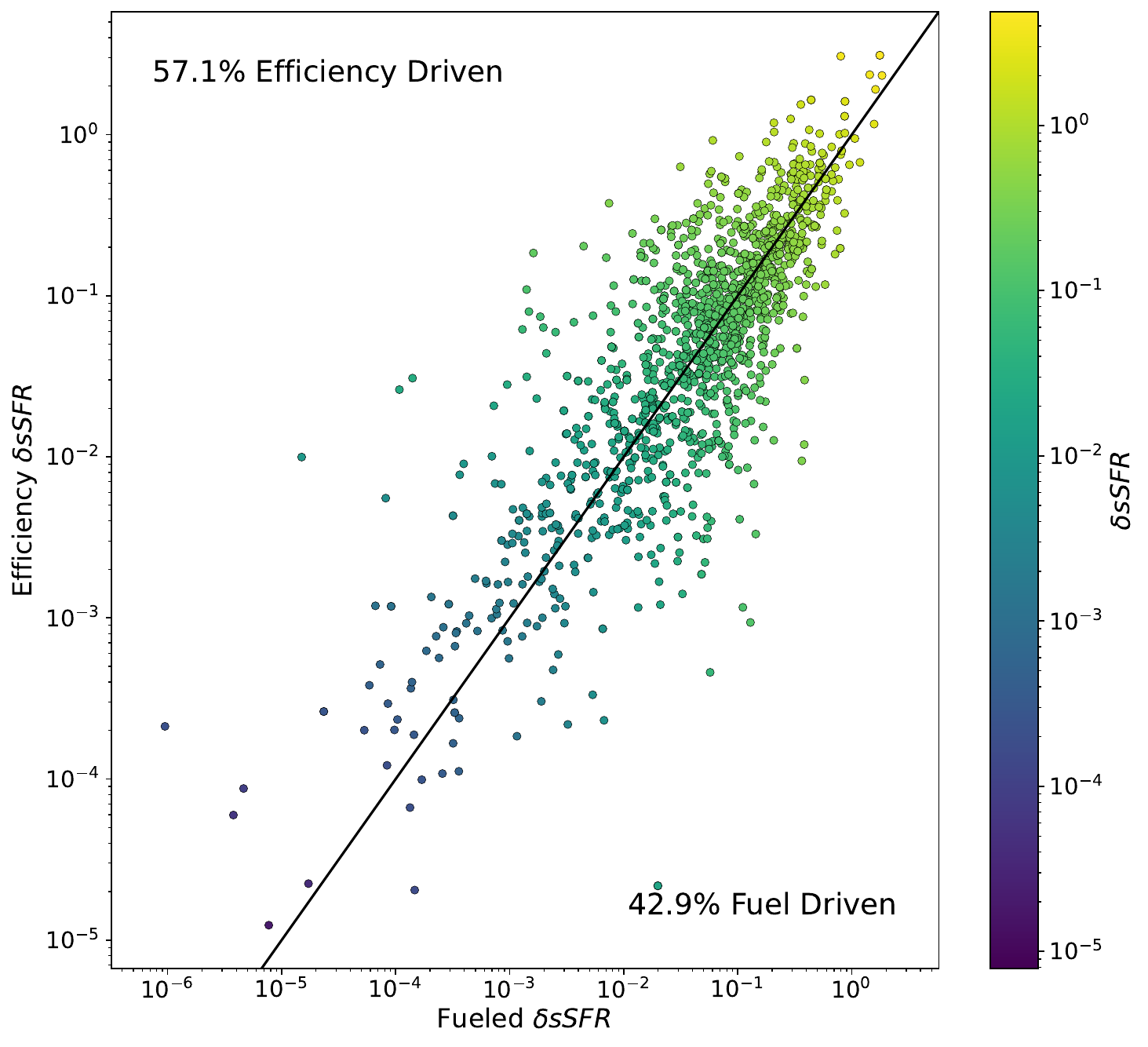}
    \caption[Comparing mean sSFR separation for host galaxies with r\textsubscript{peri} = 0-50 kpc]{Scatter plot showing the fuelled fraction of $\rm \delta sSFR$ versus the efficiency fraction of $\rm \delta sSFR$ for individual galaxies at the time of average peak post-pericentre central sSFR enhancement from Figure \ref{fig:fuelfrac}. The black diagonal line represents where fuelled $\rm \delta sSFR$ = efficiency $\rm \delta sSFR$. Points are coloured by $\rm \delta sSFR$. Some very low $\rm \delta sSFR$ galaxies are plotted but not shown.}
    \label{fig:scatterfuelfrac}
\end{figure}

To better understand the role of $\rm SFE_{H}$ in driving star formation, we now examine the individual galaxies that contribute to the average at and around the peak of $\delta \rm sSFR / \rm sSFR_{\rm i}$ in panel (a) of Figure \ref{fig:fuelfrac} (i.e. those galaxies which fall within and around the bin in which the maximum average value is calculated). 
Figure \ref{fig:scatterfuelfrac} presents a scatter plot of the efficiency $\delta \rm sSFR$ versus the fuelled $\delta \rm sSFR$ for these galaxies, with points coloured according to $\delta \rm sSFR$. 
It is important to note that, in this case, these quantities are calculated for each galaxy relative to the initial value of its predecessor at the start of their individual encounters. 
This approach differs from Figure \ref{fig:fuelfrac} and previous analyses in this work (Figures \ref{fig:sSFRHalf}, \ref{fig:propshalf1}, and \ref{fig:propsouter}), which use a moving average and calculate enhancements based on the average initial value.

From Figure \ref{fig:scatterfuelfrac}, we find that in 57.1 per cent of these galaxies, most of the enhanced star formation is driven by an enhancement in $\rm SFE_{H}$, while in 42.9 per cent of the galaxies, it is driven by an enhancement in $\rm f_{gas}$. This generally agrees with our conclusion from Figure \ref{fig:fuelfrac}, where $\rm SFE_{H}$ plays a more dominant role than gas supply in driving changes in sSFR. 
Galaxies with the largest changes in sSFR ($\delta \rm sSFR$) are located in the upper right of the plot, with the majority of these galaxies lying above the diagonal line, indicating they are classified as efficiency-driven. 
This analysis gives equal weight to each galaxy, regardless of the size of the enhancement in $\delta \rm sSFR / \rm sSFR_{\rm i}$.
Conversely, the statistical enhancements depicted in Figure \ref{fig:fuelfrac} give more weight to the galaxies with the largest increases in sSFR. 
Thus, although both analyses show that efficiency is the dominant factor, the difference in methodology explains the difference between these two metrics.

\section{Discussion}\label{sec:disc}
\subsection{Enhancements due to Close Interactions}\label{sec:sfrdisc}
We have shown that, on average in a cosmological context, sSFR is most strongly enhanced in galaxies within 0.1 Gyr after the pericentre of a close encounter with a companion galaxy (Section \ref{subsec:galprop}).
These results provide a bridge between observational pair catalog studies of interacting galaxies and high-resolution merger simulations. 

Our results are largely consistent with studies of galaxy pairs which find that SFR is enhanced for similar-mass galaxies in close pairs \citep{Ellison2008, Li2008, Scudder2012a, Patton2013, Pan2018, Bustamante2020}. 
\citet{Patton2013}, in their examination of a large sample of SDSS star-forming galaxies, found SFR enhancements as high as 2.5 times for galaxy pairs at separations within 20 kpc when compared to a control sample matched in stellar mass, redshift, local density, and environment. 
Various studies observed that such enhanced SFR is typically centrally peaked (e.g. \citealp{Scudder2012a}, \citealp{Ellison2013}, \citealp{BarreraBallesteros2015}, \citealp{Pan2019}).  
These results are consistent with our largest sSFR enhancements being found within R\textsubscript{1/2} for the 0-50 kpc r\textsubscript{peri} bin, as shown in Section \ref{subsec:galprop}.
However, we record a smaller maximum enhancement of $1.6 \pm 0.1$ times the initial pre-encounter value; this may be explained by our enhancements being calculated relative to the average sSFR before the encounter rather than to a set of control galaxies.
Additionally, our requirement that encounters be distinct from one another means that we exclude coalescing pairs, which typically have the largest SFR enhancements \citep{Ellison2013, Thorp2024}.
\citet{Patton2013} found SFR enhancements to be statistically significant out to pair separations of 150 kpc. 
In the upper panel of Figure \ref{fig:sSFRrpericompare}, we plot the average separation of the pair galaxies over the course of an encounter; we see that at the time of maximum sSFR enhancement ($\Delta t\sim$\,0.12 Gyr), galaxy pairs in the 0-50 r\textsubscript{peri} bin are, on average, at separations of 75 kpc. 
At around $\Delta$t $\approx$ 0.5 Gyr, the time when the average sSFR of our galaxies falls below their average pre-encounter value, pairs are typically at separations of 200 kpc. 
Therefore, due to the time delay with which the interaction triggers the maximum central enhancement, it is conceivable that SFR enhancements could be observed in galaxies at larger separations than would typically be classified as a close pair, as also noted in \citet{Patton2013}.
However, it should be noted that the 3D separations we utilise are not directly comparable to the projected separations used by observational studies.
\citet{Patton2020} found that projection effects can notably dilute measured sSFR enhancements.

The time delay between the pericentre of the encounter and the time at which the sSFR enhancement is maximal has significant implications for the interpretation of pair catalog studies.
These studies typically consider galaxies with a projected separation of 0-30 kpc to be closely interacting \citep{Ellison2008, Patton2013}.
\citet{Patton2020} suggest that sSFR enhancements observed at large projected separations may result from close encounters whose peak sSFR occurs at greater distances. 
This aligns with our findings, as shown in Figure \ref{fig:sSFRrpericompare}, where we observe the maximum sSFR enhancement at a mean separation of approximately 75 kpc.
Finally, \citet{ParkGottChoi2008} found that galaxies which have likely had an interaction with their companion and are now well separated are more likely to display the properties of quenched galaxies. 
This is consistent with our result shown in panel (a) of Figure \ref{fig:propshalf1}, where at $\Delta t = 1$ Gyr, the sSFR of the 0-50 kpc $r_{\text{peri}}$ bin has dropped below that of the 200-500 kpc $r_{\text{peri}}$ bin, with the latter being roughly analogous to passive evolution.

Here, through our orbital reconstruction using interpolation and encounter selection, we are able to directly link the observed enhancements in sSFR to close pericentre encounters rather than indirectly through a correlation with pair separation. 
Additionally, we have done this for a large sample size of star-forming galaxies, in a cosmological context, demonstrating that these effects are significant and occur on average.

While our study focuses on capturing the average response of galaxy properties to pericentre encounters across a broad sample, we acknowledge that factors such as mass, mass ratio, relative velocity, orbital spin, eccentricity, redshift, environment, and initial gas fraction likely influence the strength and timescale of these responses. 
Moreover, galaxies may respond differently to close encounters if they have previously experienced one \citep{Lotz2008, Torrey2012}; we refer the interested reader to Appendix \ref{appendix} for a comparison of first-passage versus all-passage trends.
Future work will aim to investigate how these parameters, as well as encounter history, modulate the effects of close interactions.

\subsection{Gas Inflows}\label{sec:gasinflowdisc}
In panel (a) of Figure \ref{fig:propshalf1}, panel (b) of Figure \ref{fig:propsouter}, and Figure \ref{fig:metal}, we find persuasive evidence suggesting an inflow of pristine gas towards the centre of a host galaxy following a close pericentric passage with its companion. 
In panel (c) of Figure \ref{fig:propshalf1}, within the 0 - 50 kpc r\textsubscript{peri} bin, we see an enhancement in $\rm f_{gas}$ immediately following the encounter at $\Delta$t = 0 Gyr in the outer shell of the host galaxy.
In both spatial regions, we observe that periods of maximum enhancements in $\rm f_{gas}$ coincide with corresponding minima in metallicity.
The enhancement of $\rm f_{gas}$ in the outer shell is followed by a more significant enhancement in $\rm f_{gas}$ within the stellar half-mass radius at a later $\Delta$t.
This further suggests an inflow of pristine, low metallicity gas.

Close SDSS galaxy pairs similarly show that interacting galaxies have lower central gas metallicities \citep{Scudder2012a, Kaneko2013, Pan2019, Bustamante2020}.
Modern high-resolution merger simulations \citep{Iono2004, Moreno2015, Blumenthal2018, Moreno2021} show that interactions trigger inflows of cold, dense gas toward the galactic centre. 
\citet{Torrey2012} specifically found that interactions transport low metallicity gas to the galactic centre, a process which is modulated by enrichment and outflows from stellar feedback. 
\citet{Torrey2012} found that the degree of this metallicity dilution is dependent on the progenitor gas content, with some gas-rich pairs even resulting in central metallicity enrichment rather than dilution.
\citet{Blumenthal2018} demonstrated that hydrodynamical shocks from a close encounter create gas filaments which then collapse into massive, dense clumps.
These clumps of dense gas then spiral towards the galactic centre as they lose angular momentum. 
Additionally, \citet{Vasiliev2022}, in their N-body merger simulations, found that interacting galaxies exert a torque on each other at the pericentre of their encounters.
Following the pericentre, the restoring force of the galaxy resuming its regular rotation triggers gas inflows towards the galactic centre. 
By tracking $\rm f_{gas}$ and gas metallicity in time relative to pericentre, we are able to demonstrate that the centrally enhanced, low metallicity gas seen in galaxy pairs observationally is very likely due to gas inflows triggered by the pericentres of close encounters. 

However, it has been suggested that the formation and motion of these dense clouds of gas may be an artefact of the inability of simulations to adequately resolve the ISM \citep{Teyssier2010}. 
Testing the sensitivity of our results to mass resolution, using the higher-resolution TNG50 run of IllustrisTNG, is an avenue for future work.
One could more directly investigate inflows in IllustrisTNG by analysing the motion and transport of gas using particle data; however, that is beyond the scope of this project.

Our analysis tracks the total hydrogen gas mass ($\rm M_{H}$), including all phases, due to the limitations of the TNG100-1 data release. However, \citet{Diemer2018} and \citet{Diemer2019} estimate the abundance of atomic and molecular hydrogen for galaxies in TNG100-1. 
For galaxies with stellar masses of ($10^{10}M_{\odot}$) to ($10^{12}M_{\odot}$) at $z = 0$, they found that molecular hydrogen accounts for roughly 5--20 per cent of the combined atomic and molecular hydrogen mass, with higher concentrations in the central regions. 
Similarly, \citet{Moreno2019} found that cold dense gas constitutes approximately 7--10 per cent of the total gas mass at pericentre in their merger simulations. 
These findings suggest that a relatively small fraction of the tracked hydrogen gas is likely star-forming gas. 
A greater portion of $\rm f_{{gas}}$ within $R_{1/2}$ is likely to be molecular gas, whereas the outer region ($R > R_{1/2}$) is more likely dominated by hot ionised gas \citep{Diemer2019}.

Future work could analyse particle-level data to separate gas phases more explicitly. However, the current methodology focuses on capturing broader trends in gas content and star formation efficiency across a large sample of galaxies in a cosmological context.

\subsection{Fuelling vs. Efficiency}\label{sec:fuelfracdisc}
One of the central goals of this study is to determine whether enhanced fuel or enhanced efficiency is the primary driver for new star formation following close encounters. 
The extent to which fuelling and efficiency contribute to driving interaction-triggered star formation is currently a subject of debate.
\citet{Pan2018} found that gas enhancement is significantly more important than efficiency in driving star formation in close pairs. 
Notably, they found efficiency to only be significant in galaxy pairs with mass ratios similar to our sample, but still much less important than gas enhancement.
On the other hand, \citet{Violino2018} found that molecular gas depletion times are shorter in pairs than in isolated galaxies matched in SFR, suggesting that galaxy pairs form stars more efficiently.
In a sample of non-interacting ALMa-QUEST galaxies, \citet{Ellison2020} found that central starbursts are driven by enhanced efficiency.
\citet{Moreno2021} found that a larger proportion of galaxies in their simulation have SFR enhancements driven by an enhanced gas content. 
\citet{Moreno2021} also found that there is a greater enhancement of $\rm M_{H_{2}}$ and SFR in the companion than in the host and that SFE is enhanced in the companion but suppressed in the host. 
\citet{Thorp2022}, in their study of 31 interacting and merging ALMa-QUEST galaxies, found that one-third of their sample's star formation is driven by efficiency, another third by fuel supply, and the remaining third by a combination of both. 
Each of these thirds displays distinct star formation histories and does not appear to correlate with interaction stage.
Direct comparison of these results to our work is complicated by varying methodologies and definitions of gas fraction and SFE. 
As discussed in Section \ref{subsec:galprop}, we define $\rm f_{gas}$ as the fraction of hydrogen gas mass to stellar mass and $\rm SFE_{H}$ as the star formation efficiency of hydrogen gas. 
However, the more standard definitions of both of these quantities specifically use molecular hydrogen gas.
Additionally, our methodology splits the contribution of fuelling and efficiency using the statistical averages of sSFR and $\rm f_{gas}$ for large numbers of galaxies as opposed to a galaxy-by-galaxy (as in \citealp{Moreno2021}) or a spaxel-by-spaxel approach (as in \citealp{Thorp2022}).

Figure \ref{fig:fuelfrac} plots the fraction of sSFR enhancement driven by enhanced $\rm f_{gas}$ and the fraction driven by enhanced efficiency.
Overall, we find that enhancements in sSFR, in both the central R\textsubscript{1/2} and outer shell of the galaxy, are primarily driven by enhanced $\rm SFE_{H}$.
The enhancement in $\rm SFE_{H}$ accounts for 70 per cent of the maximum enhancement in sSFR within R\textsubscript{1/2} and 90 per cent of the maximum enhancement in sSFR within the outer shell. 
Additionally, Figure \ref{fig:fuelfrac} shows that the $\rm SFE_{H}$ enhancement appears to sustain the sSFR as the $\rm f_{gas}$ enhancement wanes. 
The total enhancement in star formation appears to be sustained by some combination of the enhanced efficiency with which gas is converted to stars and the assembly of star-forming regions triggered by the gas inflow, both of which are accounted for in $\rm SFE_{H}$. 
Star formation is ultimately suppressed by diminishing fuel despite $\rm SFE_{H}$ remaining high, which suggests both that efficiency remains high and/or that the remaining fuelling is in a form conducive to continuing star formation.
From Figure \ref{fig:scatterfuelfrac}, using methodology more similar to \citet{Moreno2021} and \citet{Thorp2022}, we find that 57 per cent of galaxies at the peak sSFR enhancement are individually efficiency driven. 
Galaxies with the largest change in sSFR from their individual initial values appeared to be more likely to be efficiency driven, contributing to the stark result in Figure \ref{fig:fuelfrac}.
These results come with the caveat that our measurement of efficiency includes both the efficiency of converting gas into stars, but also the efficiency of assembly in star-forming regions, i.e. the fraction of cold-dense gas to total gas (Eqns. \ref{eq:SFEHeq} and \ref{eq:fineq}).
A likely mechanism is that a close encounter triggers an inflow of pristine, low metallicity gas, from the outskirts of the galaxy and the circumgalactic medium, towards the centre of the galaxy, which enhances the available fuel for star formation. 
The total mass of this gas relative to the stellar mass is captured in our measurement of $\rm f_{gas}$.
However, this inflow also assembles regions of cold-dense molecular gas (i.e. hydrogen gas becomes cooler and denser) from which stars are able to form more efficiently, which is captured in our measurement of $\rm SFE_{H}$.
Additionally, the inflow process may cause the gas to change state from atomic to molecular hydrogen gas \citep{Kaneko2017}. 

\citet{Kaneko2017} found that external shocks due to an interaction assemble molecular gas and trigger star formation. 
They found that during inflow, gas may cool and condense or atomic gas may collide with the surface of giant molecular clouds and condense. 
\citet{Moreno2019}, in their high-resolution merger simulations, similarly found a mass exchange pipeline whereby warm gas is cooled and then condensed to cold, dense molecular gas.
However, at each iteration some gas is lost as hot gas which cannot be cooled again in the required timescale. 
\citet{Kaneko2017} also found that diffuse gas collisions with giant molecular clouds, as well as collisions between molecular clouds, can act to destabilise them and trigger star formation. 
These mechanisms would result in higher $\rm SFE_{H}$ both through the enhanced efficiency with which gas is converted to stars, but also with the assembly of cold, dense clouds of molecular gas. 
As such, we conclude that gas inflows are likely the primary mechanism responsible for triggering star formation in our sample of simulated interacting galaxies.
However, this likely occurs not by augmenting the fuel supply, but by inducing higher SFE through the mechanisms discussed above.

\section{Conclusions}\label{sec:conclusion}

In this study, we have aimed to more directly link the changes in a galaxy's star-forming properties to the pericentre of its encounter with a close companion galaxy for a large sample of massive galaxies in a cosmological context. 
We construct a dataset of galaxy-galaxy encounters containing discrete pericentres by utilizing a sample of IllustrisTNG100-1 galaxy pairs \citep{Patton2020} whose interacting orbits have been reconstructed through 6D kinematic interpolation \citep{Patton2024}.
We use this dataset to track the evolution of galaxies throughout the encounter.
For galaxies which have an encounter with their companion with an r\textsubscript{peri} of $\leq 50$ kpc, we find the following:
\begin{itemize}
    \item The sSFR within the central region of the galaxy (one stellar half-mass radius) is enhanced on average by a factor of $1.6 \pm 0.1$ times its initial value following pericentre passage (Figure \ref{fig:sSFRHalf}).
    \item The maximum sSFR enhancement occurs approximately 0.1 Gyr after the encounter, where galaxy pairs are at average separations of 75 kpc (Figure \ref{fig:sSFRrpericompare}).
    \item The $\rm f_{gas}$ and $\rm SFE_{H}$ are enhanced on average by factors of \( 1.2 \pm 0.1 \) and \( 1.4 \pm 0.1 \), respectively, on similar timescales (Figure \ref{fig:propshalf1}).
    \item In the outer shell of the galaxy (beyond one stellar half-mass radius), sSFR, $\rm f_{gas}$, and $\rm SFE_{H}$ enhancements are smaller and reach their maximum approximately 0.05 Gyr earlier than in the central region (Figure \ref{fig:propsouter}).
    \item Significant gas metallicity dilution follows the pericentre, with a smaller dilution in the outer shell and significant, long-lived dilution in the central region (Figure \ref{fig:metal}).
\end{itemize}

A central objective of this study was to determine whether the enhanced star formation rates in interacting galaxies are best explained by an increased fuel supply or an enhanced efficiency at which star formation occurs and star-forming regions are established. 
To this purpose, we defined the fuelled $\delta \rm sSFR / sSFR_{\rm i}$ and efficiency $\delta \rm sSFR / sSFR_{\rm i}$, which describe the relative contributions of the enhancements in $\rm f_{gas}$ and SFE\textsubscript{H} to the enhancement in sSFR.
Significantly, we found that enhancements in sSFR as a result of the pericentric passage of a companion are primarily driven by enhancements in SFE\textsubscript{H}. 
More specifically, we found:
\begin{itemize}
    \item Within the central region, 70 per cent of the maximum enhancement in sSFR can be attributed to the enhancement in star formation efficiency and the efficiency at which star-forming regions are created (Figure \ref{fig:fuelfrachalf}). 
    \item In the outer shell, less than 15 per cent of the maximum enhancement in sSFR can be attributed to an enhanced availability of fuel (Figure \ref{fig:fuelfracring}). 
\end{itemize}

These findings are consistent with a model wherein the pericentric passage of a galaxy's companion initiates shocks within the galaxy. This could occur through mechanisms such as hydrodynamical forces or torques due to the spin-orbit orientation of the galaxy pair, inducing an inflow of pristine gas towards the galaxy centre.
We discuss how this inflow may trigger the condensing of gas to create new dense regions of gas, the conversion of atomic gas to molecular gas, the collision of giant molecular clouds, or trigger instabilities within these clouds, allowing for easier collapse. 
These mechanisms would all be accounted for in our measurement of SFE\textsubscript{H} and would result in enhanced and sustained star formation.

\section*{Acknowledgements}

We thank the anonymous referee for a thoughtful and constructive referee’s report.
We thank all members of the IllustrisTNG collaboration for making their data available. 
DRP, SC and SE gratefully acknowledge NSERC of Canada for Discovery Grants which helped to fund this research.
The IllustrisTNG simulations were undertaken with compute time awarded by the Gauss Centre for Super- computing (GCS) under GCS Large-Scale Projects GCS-ILLU and GCS-DWAR on the GCS share of the supercomputer Hazel Hen at the High Performance Computing Center Stuttgart (HLRS), as well as on the machines of the Max Planck Computing and Data Facility (MPCDF) in Garching, Germany. LF gratefully acknowledges support from Queen’s University. Nikhil Arora and Kristine Spekkens are thanked for insightful discussion and useful suggestions.
\section*{Data Availability}
The IllustrisTNG data used in this work is publicly available at \url{http://www.tng-project.org}.



\bibliographystyle{mnras}
\bibliography{References} 




\appendix

\section{Influence of First Passages on Results}\label{appendix}

We have aimed to reduce the cumulative effects of multiple pericentric passages on our results by applying the encounter selection criteria in Section \ref{subsec:encount} that ensure pericentres are sufficiently spaced in time, allowing any effects from prior passages to dissipate before the next one. To investigate whether this has had an impact on our results, we defined a subset of encounters corresponding to only encounters experiencing their first pericentric passages within our redshift range $0 \leq z < 1$. This first-passage subset comprises 70 per cent of the total dataset.
As discussed in Section 2.2, it is important to note this does not account for potential earlier passages outside this redshift range.

\begin{figure}
    \centering
    \includegraphics[width=\columnwidth]{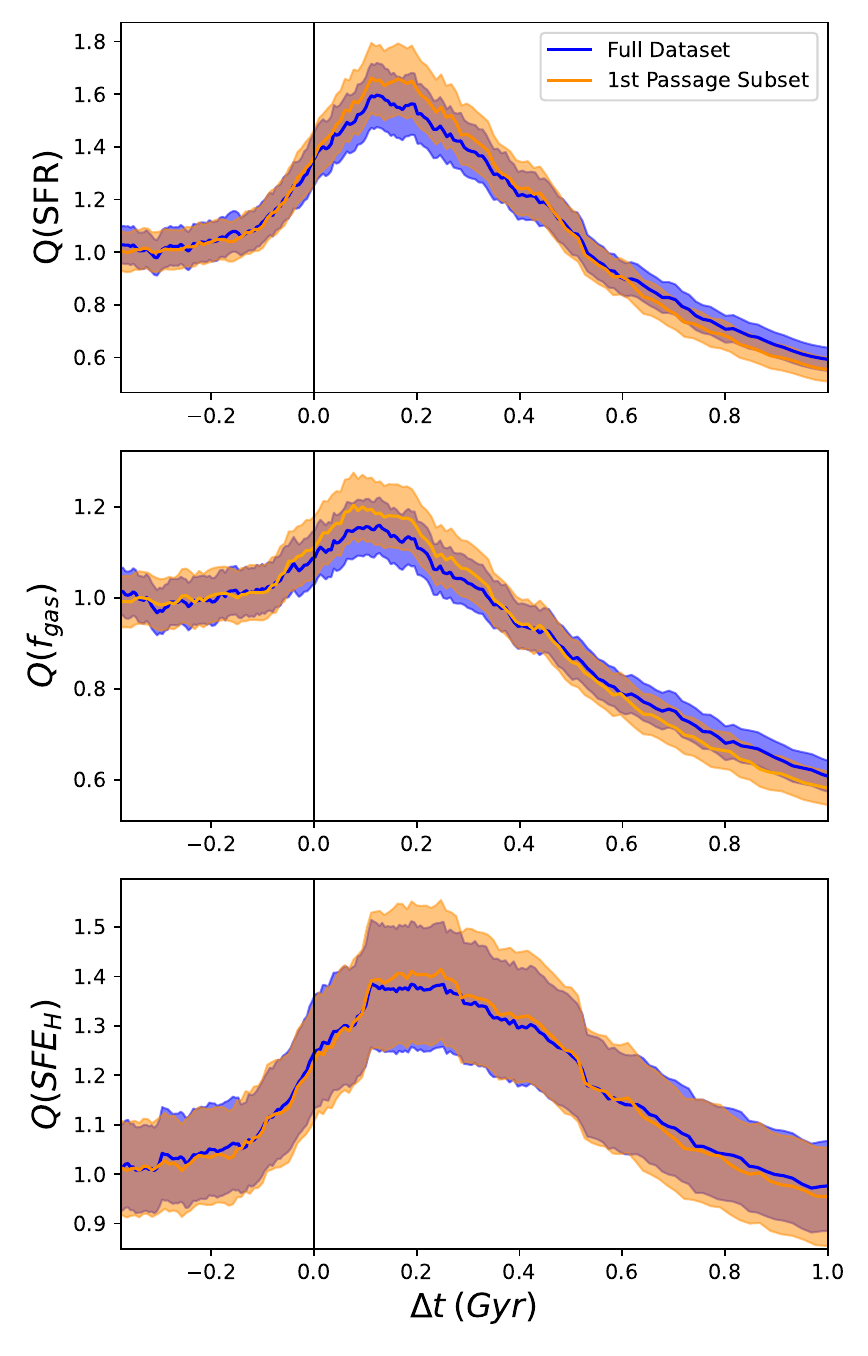}
    \caption{Comparison of mean enhancements in the galaxy properties of interest versus time relative to the pericentre, measured within R\textsubscript{1/2}, for all host galaxies (blue) and for the subset limited to first-passage encounters (orange), both with r\textsubscript{peri} = 0–50 kpc. 
    }
    \label{fig:firstpasscompare}
\end{figure}

In Figure \ref{fig:firstpasscompare}, we compared the maximum enhancement in sSFR, f\textsubscript{gas} and SFE\textsubscript{H} for host galaxies in the 0-50 kpc r\textsubscript{peri} bin of interest for the first-passage subset and the full dataset. We find that these enhancements generally follow the same trend and the difference between the maximum enhancements in each quantity is well within the error. Additionally we find the timing of the maximum enhancement in each quantity to be consistent between the full dataset and the first passage subset. 



\bsp	
\label{lastpage}
\end{document}